\documentclass[a4paper,11pt]{article}
\pdfoutput=1

\usepackage{jheppub}
\usepackage{natbib}
\usepackage{braket}

\usepackage[utf8]{inputenc}
\usepackage{todonotes}
\usepackage{amssymb,amsmath,amsbsy}
\usepackage{braket}
\usepackage{graphicx,color}
\usepackage{hyperref}
\newcommand{\mA}{\mathcal{A}}
\newcommand{\mB}{\mathcal{B}}

\newcommand{\p}{\partial}

\newcommand{\mI}{\mathbb{I}}
\newcommand{\mH}{\mathcal{H}}
\newcommand{\nn}{\nonumber}
\newcommand{\lb}{\left(}
\newcommand{\rb}{\right)}
\newcommand{\ee}{\end{equation}}
\newcommand{\bea}{\begin{eqnarray}}
\newcommand{\eea}{\end{eqnarray}}

\newcommand{\tr}{\text{tr}}

\title{Monotonic multi-state quantum $f$-divergences}

 \author{Keiichiro Furuya$^a$,}
 \author{Nima Lashkari$^{a,b}$}
 \author{and Shoy Ouseph$^b$}

 \affiliation[a]{Department of Physics and Astronomy, Purdue University, West Lafayette, IN
 47907, USA}
\affiliation[b]{School of Natural Sciences, Institute for Advanced Study, Princeton, New Jersey 08540, USA}

\emailAdd{kfuruya@purdue.edu}
 \emailAdd{nima@purdue.edu}
 \emailAdd{souseph@purdue.edu}

\abstract{We use the Tomita-Takesaki modular theory and the Kubo-Ando operator mean to write down a large class of multi-state quantum $f$-divergences and prove that they satisfy the data processing inequality. For two states, this class includes the $(\alpha,z)$-R\'enyi divergences, the $f$-divergences of Petz, and the measures in \cite{matsumoto2015new} as special cases. The method used is the interpolation theory of non-commutative $L^p_\omega$ spaces and the result applies to general von Neumann algebras including the local algebra of quantum field theory. We conjecture that these multi-state R\'enyi divergences have operational interpretations in terms of the optimal error probabilities in asymmetric multi-state quantum state discrimination.}

\makeatletter
\gdef\@fpheader{}
\makeatother
\begin{document}

\maketitle

\section{Introduction}

\paragraph{Motivation:} In classical physics, the state of a system is a probability distribution $p(x)$ over the configuration space $X$. 
To distinguish different states one needs to compare probability distributions.
The Kullback-Leibler divergence 
\begin{eqnarray}
 D_{KL}(\{q\}\|\{p\})=\sum_{x\in X} q(x)\log (q(x)/p(x))
\end{eqnarray}
is a distinguishability measure that plays a central role in information theory and has an interpretation in terms of the thermodynamic free energy difference of the state $\{q\}$ from the equilibrium distribution $\{p\}$ \cite{vedral2002role}. It is non-negative, non-degenerate\footnote{It is zero if and only if the probability measures are the same.} and
monotonically non-increasing under the action of a classical channel.\footnote{A classical channel a stochastic map $T:X \to Y$ with $\sum_{y\in Y}T(y|x)=1$. In other words, a classical channel is a conditional probability distribution.} The thermodynamic interpretation of relative entropy explains why this measure of distinguishability is not symmetric under the exchange of $\{q\}$ and $\{p\}$. The monotonicity under classical channels is an essential property that any reasonable distinguishability measure should satisfy.\footnote{Intuitively, this is because either the channel is noiseless in which case the distinguishability remains the same, or it is noisy and the distinguishability decreases.} We say a quantity satisfies the data processing inequality if it is monotonic under the action of a channel. One can consider symmetric distinguishability measures such as the log-fidelity
\begin{eqnarray}
D_{1/2}(\{q\},\{p\})=-2\log\sum_{x\in X} \sqrt{q(x)} \sqrt{p(x)}
\end{eqnarray}
or, in general, a one-parameter family $\theta\in (0,1)$ of non-negative, non-degenerate measures 
\begin{eqnarray}\label{Renyiclassical}
D_\theta(\{q\}\|\{p\})= \frac{1}{(\theta-1)}\log\sum_{x\in X}q(x)^\theta p(x)^{1-\theta}
\end{eqnarray}
that interpolate between $D_{KL}(\{q\}\|\{p\})$ at $\theta=1$ and log-fidelity at $\theta=1/2$ and satisfy the data processing inequality.
% \footnote{Up to an overall factor these measures are the Renyi divergences of the two probability distribution.}
It is tempting to generalize to a multi-state measure 
\begin{eqnarray}\label{generating}
 &&D_{\theta_1,\cdots,\theta_n}(\{p_1\},\cdots, \{p_n\})=\frac{-1}{(1-\theta_1)\cdots (1-\theta_n)}\log\lb \sum_{x\in X}p_1(x)^{\theta_1}\cdots p_n(x)^{\theta_n}\rb\nn\\
 &&\theta_1+\cdots +\theta_n=1
\end{eqnarray}
as a functional that interpolates between $D_{KL}(\{p_i\}\|\{p_j\})$ and their corresponding log-fidelities for different $i$ and $j$. Note that the parameters $(\theta_1,\cdots, \theta_n)$ can be thought of as a probability distribution. We are not aware of any arguments in the literature that proves that the measure above satisfies the data processing inequality.
In this work, we write down a quantum generalization of the above measure and prove that it satisfies the data processing inequality.

In quantum mechanics, the state of a system is a completely positive (CP) map from the algebra of observables to complex numbers $\omega:\mA\to \mathbb{C}$ with $\omega(1)=1$. If the observable algebra is the algebra of $d\times d$ complex matrices a state is a density matrix (positive operator with unit trace): $\omega>0$ with $\tr(\omega)=1$. The quantum relative entropy
\begin{eqnarray}
 S(\psi\|\omega)=\tr(\psi\log\psi)-\tr(\psi\log\omega)
\end{eqnarray}
is a measure of distinguishability of the density matrix $\psi$  from $\omega$. It is non-negative, non-degenerate and has an operational interpretation in asymptotic asymmetric hypothesis testing \cite{hiai1991proper}. One can define a symmetric distinguishability measure called log-fidelity:
\begin{eqnarray}
D_{1/2}(\psi\|\omega)=-2\log \tr\sqrt{\omega^{1/2}\psi \omega^{1/2}}\ .
\end{eqnarray}
Since in quantum mechanics the density matrices need not commute there can be many non-commutative versions of the R\'enyi divergences in (\ref{Renyiclassical}) that interpolate between the relative entropy $S(\psi\|\omega)$ and log-fidelity. Two important families of measures of this kind are the {\it Petz R\'enyi divergences} and the {\it sandwiched R\'enyi divergences}, respectively
\begin{eqnarray}\label{Petzandsandwiched}
 &&D_\theta(\psi\|\omega)=\frac{1}{\theta-1}\log \tr\lb \psi^\theta \omega^{1-\theta}\rb\nn\\
 &&S_\theta(\psi\|\omega)=\frac{1}{\theta-1}\log\tr\lb \lb \omega^{\frac{1-\theta}{2\theta}}\psi \omega^{\frac{1-\theta}{2\theta}}\rb^{\theta} \rb\ .
\end{eqnarray}
These two families are distinguished because they satisfy the data processing inequality. They have operational interpretations in  hypothesis testing \cite{mosonyi2015quantum}. A larger two-parameter family of R\'enyi divergences called $(\alpha,z)$-{\it R\'enyi relative entropy} interpolates between the two families \cite{audenaert2015alpha}. In our notation, we call them the $(\theta,r)$-R\'enyi divergences
\begin{eqnarray}\label{alphazintro}
S_{\theta,r}(\psi\|\omega)
=\frac{1}{\theta-1}\log\tr\left[ \lb \omega^{\frac{1-\theta}{2r}}\psi^{\frac{\theta}{r}}\omega^{\frac{1-\theta}{2r}}\rb^r\right]\ .
\end{eqnarray}
In fact, they were introduced earlier in \cite{jaksic2011entropic} as entropic measures in out-of-equilibrium statistical mechanics. They satisfy the data processing inequality in the range of $(\theta,r)$ specified in \cite{zhang2020wigner}.

The generalization of hypothesis testing to a multi-state setup is often called quantum state discrimination. In the asymmetric case, we are given some state and the task is to identify whether the state is $\omega$ or any of the alternative hypotheses $\psi_1, \cdots, \psi_k$ by performing measurements on infinite number of copies of $\omega$. The distinguishability measure with a natural operational interpretation in this case is \cite{brandao2010generalization}
\begin{eqnarray}
 &&\min_{\psi\in K}S(\psi\|\omega)\qquad K=\{\psi_1,\cdots, \psi_k\}\ .
\end{eqnarray}
Motivated by quantum state discrimination, in this work, we introduce a large family of multi-state quantum R\'enyi divergences that interpolate between various $S(\psi_i\|\omega)$ and satisfy the data processing inequality. We generalize our measures to multi-state quantum $f$-divergences.

\paragraph{Method:} 
We employ three main tools to construct the multi-state R\'enyi divergences and prove their monotonicity. The first tool is the Araki-Masuda non-commutative $L^p_\omega$ spaces \cite{araki1982positive} that we review in section \ref{sec:Lp} and $\ref{sec:Lprho}$. In particular, we use the Riesz-Thorin theorem to prove that a contraction operator $F$ does not increase the $L^p_\omega$ norm of the vectors.
The second tool is the monotonicity of the relative modular operator in the Tomita-Takesaki modular theory. A quantum channel $\Phi^*$ corresponds to a contraction $F$ in the GNS Hilbert space. The relative modular operator satisfies the inequality
\begin{eqnarray}
 F^\dagger \Delta_{\psi|\omega}F\leq \Delta_{\Phi^*(\psi)|\Phi^*(\omega)}\ .
\end{eqnarray}
The third tool is the Kubo-Ando operator mean for positive operators $X$ and $Y$:
\begin{eqnarray}
 X\sharp_f Y=X^{1/2}f(X^{-1/2}Y X^{-1/2})X^{1/2}
\end{eqnarray}
and an operator monotone function $f$ with $f(1)=1$. The Kubo-Ando mean has the property that if $X_1\leq Y_1$ and $X_2\leq Y_2$ then 
\begin{eqnarray}
X_1\sharp_f X_2\leq Y_1\sharp_f Y_2\ . 
\end{eqnarray}
This allows us to construct multi-state operator monotonicity inequalities of the type
\begin{eqnarray}
 F^\dagger(\Delta_{\psi_1|\omega}\sharp_f\Delta_{\psi_2|\omega})F\leq
 (\Delta_{\Phi^*(\psi_1)|\Phi^*(\omega)}\sharp_f\Delta_{\Phi^*(\psi_2)|\Phi^*(\omega)})\ .
\end{eqnarray}
The $L^p_\omega$-norm of the vector $\lb\Delta_{\psi_1|\omega}\sharp_f\Delta_{\psi_2|\omega}\rb^{1/2}\ket{\omega^{1/2}}$ is the building block of the class of multi-state R\'enyi divergences we construct in this work.

\paragraph{Summary of results:}

In the case of two states, in equation (\ref{alphaz}), we write the  $(\theta,r)$-R\'enyi divergences as the $(r,\omega)$-norm of a vector in the $L^p_\omega$ spaces.\footnote{A similar expression appears in \cite{jaksic2011entropic}.} We generalize them to two-state divergences in (\ref{Sfr}).
We use the monotonicity of the relative modular operator and the Riesz-Thorin theorem (see appendix \ref{App:Riezs}) from the complex interpolation theory to prove that these two-state measures satisfy the data processing inequality in the range $r\geq 1$.\footnote{The monotonicity of $(\theta,r)$-R\'enyi divergences was shown using different methods in \cite{hiai2013concavity}.} 
% An important step in this proof is the Stinespring factorization of a unital CP map as the composition of a $*$-representation followed by a contraction. The representation map leaves the $(\theta,r)$-R\'enyi divergences unchanged, but they are non-increasing under a compression map.

Section \ref{sec:multistate} generalizes the discussion to multiple states. First, in section \ref{sec:multiineq}, we use the complex interpolation theory to prove a generalization of the H\"{o}lder inequality to von Neumann algebras. This section follows the arguments in \cite{araki1982positive}, and can be skipped by the readers who are only interested in the multi-state R\'enyi measures. Then, in section \ref{sec:threestates}, we use the Kubo-Ando geometric mean to introduce the three-state $f$-divergence in (\ref{threestateRenyi}) and prove that they are monotonically non-increasing under quantum channels. This measure depends on an arbitrary operator monotone function $f$ with $f(1)=1$, the parameters $\theta_1,\theta_2$ with $0\leq \theta_1+\theta_2\leq 1$, $r\geq 1/2$ and three states $\psi_1, \psi_2$ and $\omega$. Specializing to the case $f(x)=x^\alpha$ with $\alpha\in [0,1]$, in matrix algebras we obtain the three-state R\'enyi divergences in (\ref{threestateAlpha}).\footnote{We prove the monotonicity only in the range $r\geq 1$.} In a special case, this measure reduces to the R\'enyi measures in \cite{matsumoto2015new}:
\begin{eqnarray}
 \bar{S}_\theta(\psi\|\omega)=\frac{1}{\theta-1}\log \tr\lb\omega^{1/2}\lb\omega^{-1/2}\psi \omega^{-1/2}\rb^\theta\omega^{1/2} \rb\ .
\end{eqnarray}
We write down an $n$-state $f$-divergences in (\ref{multistateRenyif}), multi-state R\'enyi divergences in (\ref{multistateRenyialpha}) and prove that they satisfy the data processing inequality. In matrix algebras, this multi-density matrix measure is (\ref{multiRenyi}).

In section \ref{sec:infinite}, we discuss our construction in arbitrary von Neumann algebras, focusing on the case where a trace does not exist. This is important for the applications of this work to infinite dimensional quantum systems such as the algebra of local observables in Poincare-invariant quantum field theory. In section \ref{sec:discrimination}, we conjecture that similar to the Petz divergences and the sandwiched R\'enyi divergences, the multi-state R\'enyi divergences in section \ref{sec:multistate} have operational interpretations in terms of the optimal error probabilities in various quantum state discrimination setups.

For the marginals of multi-partite systems, one can introduce the so-called swiveled R\'enyi measures \cite{Wilde_2015,dupuis2016swiveled,berta2015renyi}. In the case all $a_S$ in swiveled measures are non-negative they can be understood as a special case of the multi-state measures introduced in this work.

\section{Operator \texorpdfstring{$L^p$}{} spaces}\label{sec:Lp}

This section reviews the construction of the operator $L^p$ spaces in finite dimensional matrix algebras.
The observable algebra of a $d$-level quantum system is the algebra $\mA$ of $d\times d$ complex matrices. The linear map
\begin{eqnarray}
   &&\mA\ni a\to \ket{a}=(a\otimes \mI)\ket{e}\nn\\
   &&\ket{e}=\sum_k \ket{k,k}
\end{eqnarray}
represents the algebra on a Hilbert space $\mH_e$ with the inner product 
\begin{eqnarray}
\braket{a_1|a_2}=\tr(a_1^\dagger a_2)\ .
\end{eqnarray}
We use the simplified notation
\begin{eqnarray}
   &&a\ket{e}\equiv (a\otimes \mI)\ket{e}\nn\\
   &&a'\ket{e}\equiv (\mI\otimes a')\ket{e}
\end{eqnarray}
and refer to the algebra of operators $a'\equiv (\mI\otimes a')$ as $\mA'$, the commutant of $\mA$. The Hilbert space norm of a vector is
\begin{eqnarray}
  \|\ket{a}\|\equiv \|a\|_2=\tr(a^\dagger a)^{1/2}
\end{eqnarray}
and its $\infty$-norm (operator norm) is
\begin{eqnarray}
  \|\ket{a}\|_\infty\equiv  \|a\|_\infty=\sup_{\|\ket{\chi}\|=\|\ket{\Psi}\|=1}\braket{\chi|a\Psi}\ .
\end{eqnarray}
% Every operator has a left polar decomposition $a=a_+u$ with $a_+=\sqrt{aa^\dagger}$ a positive operator and $u$ a unitary. 

The advantage of the Hilbert space representation $\mH_e$ is that one can think of superoperators $\Phi:\mA\to \mA$ as linear operators $F:\mH_e\to \mH_e$:
\begin{eqnarray}
   F\ket{a}=\Phi(a)\ket{e}\ .
\end{eqnarray}
Linear maps $\Phi$ that are completely positive (CP) and unital are specially important in physics. In the Hilbert space, they are represented by operators that are {\it contractions}, i.e. $\|F\|_{\infty}\leq1$.\footnote{Consider a unital CP map $\Phi: \mA \rightarrow \mB$. Using the Stinespring dilation theorem, the map decomposes as $\Phi(a) = W^\dagger a W$ where $W$ is an isometry since $\Phi$ is unital. The action of the map on the GNS Hilbert space is given by
\begin{eqnarray}
\Phi(a)\ket{\Omega_B} = W^\dagger a \ket{\Omega_A}
\end{eqnarray}
where $W$ satisfies $W\ket{\Omega_B} = \ket{\Omega_A}$. The GNS operator $F$ corresponding to $\Phi$ is defined by $\Phi(a)\ket{\Omega_B} = Fa\ket{\Omega_A}$. Since $\mA\ket{\Omega_A}$ is dense in $\mH_A$, the corresponding GNS operator is a co-isometry $F=W^\dagger$ and a contraction.
} It is clear that $F$ can never increase the $2$-norm of vectors
\begin{eqnarray}
   \|F\ket{a}\|\leq \|\ket{a}\|\ .
\end{eqnarray}
It cannot increase the operator norm either because 
\begin{eqnarray}\label{contractinf}
   \|F\ket{a}\|_\infty&&=
%   \sup_{\ket{\Psi},\ket{\Omega}\in \mH_e} |\braket{\Psi|\Phi(a)|\Omega}|\nn\\
    \sup_{\|\ket{\Psi}\|=\|\ket{\chi}\|=1} |\braket{\Psi|F(a\otimes 1)|\chi}|\nn\\
    &&\leq \|F\|_\infty\|(a\otimes 1)\|_\infty\leq \|a\|_\infty\ .
\end{eqnarray}
The $2$-norm and the $\infty$-norm are special cases of the $p$-norms (Schatten norms) defined by
\begin{eqnarray}\label{Lpnorm}
  \forall p\in [1,\infty]:\qquad \|a\|_p=\tr(a_+^p)^{1/p}
\end{eqnarray}
where $a=a_+u$ is the left polar decomposition of $a$ in terms of the positive semi-definite operator $a_+$ and unitary $u$.
For $p\in (0,1)$, they are quasi-norms because they no longer satisfy the triangle inequality 
$\|a_1+a_2\|_p\nleq \|a_1\|_p+\|a_2\|_p$. 
The Hilbert space norm and the operator norm correspond to $p=2$ and $p=\infty$, respectively. 
Since the map between the operators $a$ and the vectors $\ket{a}$ in matrix algebras is one-to-one we define the $p$-norm of a vector in the Hilbert space to be the $p$-norm of the operator that creates it:
\begin{eqnarray}
   \|\ket{a}\|_p\equiv \|a\|_p\ .
\end{eqnarray}
Note that since $\|a\|_p=\|u a v\|_p$ for any unitary $u,v$ the $p$-norm of a vector satisfies
\begin{eqnarray}
   \|u u'\ket{a}\|_p=\|\ket{a}\|_p
\end{eqnarray}
where $u\in \mA$ and $u'\in \mA'$ are unitaries.

We define the superoperator norms\footnote{Note that, by definition, $\|T\|_{\infty} = \| T\|_{(2\to 2)}$.} 
\begin{eqnarray}\label{supernorm}
   \|\Phi\|_{(p_0\to p_1)}\equiv\sup_{a\in \mA}\frac{\|\Phi(a)\|_{p_1}}{\|a\|_{p_0}}
\end{eqnarray}
and the norm for their corresponding operators 
\begin{eqnarray}\label{supernorm2}
&&F\ket{a}=\ket{\Phi(a)}\nn\\
&&   \|F\|_{(p_0\to p_1)}\equiv\|\Phi\|_{(p_0\to p_1)}\ .
\end{eqnarray}

A complete normed vector space is called a Banach space. Since the Hilbert space norm is complete with respect to the $2$-norm
\begin{eqnarray}\label{innerGNS}
       &&\braket{a_1|a_2}=\tr(a_1^\dagger a_2)\nn\\
       &&\braket{a|a}=\|a\|_2^2,
\end{eqnarray}
we sometimes refer to the Hilbert space $\mH_e$ as the $L^2$ Banach space, or the $L^2$ space in short. By analogy, we call the algebra $\mA$ with the operator norm the $L^\infty$ space.\footnote{Note that the algebra itself is a linear vector space.} The representation $a\to \ket{a}$ is then a linear map from $L^\infty\to L^2$. We could also define the linear map $a\to e_a=\ket{a}\bra{e}$ that sends the algebra to a linear space of operators in $B(\mH_e)$ that we denote by $\mA_*$ and call the {\it predual} of $\mA$. The subspace of operators $\ket{a_+}\bra{e}$ is in one-to-one correspondence with the subspace of unnormalized pure density matrices $\ket{a_+^{1/2}}\bra{a_+^{1/2}}$ of the algebra $\mA\otimes \mA'$.
The predual $\mA_*$ equipped with the $1$-norm $\tr((e_a)_+)$ is called the $L^1$ space. Since the maps $a\to \ket{a}$ and $a\to e_a$ are bijections in matrix algebras we can think of the $L^1$, $L^2$ and $L^\infty$ spaces as the same space with different norms. As the dimension of algebra goes to infinity an operator with finite $2$-norm has finite $\infty$-norm but not necessarily a finite $1$-norm. So we have the hierarchy $L^1\subseteq L^2\subseteq L^\infty$.

Our Hilbert space inner product is a map from $L^2\times L^2\to \mathbb{C}$ that is anti-linear in the first variable. It could alternatively be interpreted as a map from $L^1\times L^\infty\to \mathbb{C}$:
\begin{eqnarray}\label{innerprodduality}
   \braket{a|b}=\tr(a^\dagger e_b)
\end{eqnarray}
where $e_b\in L^1$. An important property of an inner product is the Cauchy-Schwarz inequality:
\begin{eqnarray}
|\braket{a|b}|^2\leq \braket{a|a}\braket{b|b}\ .
\end{eqnarray}
% where 
% \begin{eqnarray}
%   \braket{a|a}^{1/2}=\tr(a^\dagger a)^{1/2}\ .
% \end{eqnarray}
% is the Hilbert space norm of the vector $\ket{a}$. 
The Cauchy-Schwarz inequality is saturated when $\ket{a}$ and $\ket{b}$ are parallel. This allows us to write
\begin{eqnarray}\label{dualL2}
   \|\ket{b}\|=\sup_{\|\ket{a}\|=1}|\braket{a|b}|\ .
\end{eqnarray}
Similarly, we can use (\ref{innerprodduality}) to write the operator norm $\|b\|_\infty$ as
\begin{eqnarray}\label{dualL1}
   \|b\|_\infty=\sup_{\tr((e_a)_+)=1}|\tr(e_a b)|\ .
\end{eqnarray}
We say the space $L^\infty$ is dual to $L^1$.

The generalization of the Cauchy-Schwarz inequality to the $L^p$ spaces is called the operator H\"{o}lder inequality 
\begin{eqnarray}\label{holder}
\forall p\in [1,\infty]:\qquad \|a^\dagger b\|_1\leq \|a\|_q\|b\|_p
\end{eqnarray}
and $1/p+1/q=1$. 
More generally, if $1/p_0+1/p_1=1/r$ with $r>1$ the operator H\"{o}lder inequality says
\begin{eqnarray}
\|a^\dagger b\|_r\leq \|a\|_{p_0}\|b\|_{p_1}\ .
\end{eqnarray}
In the range $p_0\in (0,1)$, the  parameter $p_1$ is negative and we have a reverse H\"{o}lder inequality  
\begin{eqnarray}\label{reverseHolder}
   \forall p_0\in (0,1):\qquad  \|a\|_{p_0}\|b\|_{p_1}\leq \|a^\dagger b\|_r\ .
\end{eqnarray}
The reverse H\"{o}lder inequality follows from the H\"{o}lder inequality and the property  $\|a^{-1}\|_{-p}=\|a\|_p^{-1}$ \cite{beigi2013sandwiched}.
We will prove the generalization of the operator H\"{o}lder inequality in an arbitrary von Neumann algebra in section \ref{sec:multiineq}.

We can realize the $p$-norm of the vector $\ket{a}\in \mH_e$ as an inner product between $\ket{a}$ and a vector $\ket{x_0}$ in the Hilbert space $\mH_e$:
\begin{eqnarray}\label{Holdereq}
    &&\|\ket{a}\|_p=\tr(a_+^p)^{\frac{1}{p}}=\tr(a_+^p)^{\frac{1}{p}-1}\braket{a_+^{p-1}|a_+}=\frac{\braket{a_+^{p-1}|a_+}}{\|\ket{a_+^{p-1}}\|_{q}}=\braket{x_0|a_+}\ .
\end{eqnarray}
The vector $\ket{x_0}\sim \ket{a_+^{p-1}}$ is normalized to have $\|\ket{x_0}\|_q=1$. 
It follows from the H\"{o}lder inequality that 
\begin{eqnarray}
    |\braket{b|a_+}|\leq \|b^{\dagger}a_+\|_1\leq \|a\|_p\|b\|_q\ .
\end{eqnarray}
We can absorb the unitaries in the polar decomposition of $a$ in $b$ to write 
\begin{eqnarray}
 |\braket{b|a}|\leq \|a\|_p\|b\|_q\ .
\end{eqnarray}
% Equation (\ref{Holdereq}) implies that the H\"{o}lder inequality is saturated for $a = a_+u$ where $b\sim a_+^{p-1} u$ is the left polar decomposition of $b$. 
The $p$-norm is the maximum overlap between $\ket{a}$ and the vectors in the Hilbert space that are normalized to have unit $q$-norm:
\begin{eqnarray}\label{sup}
  \forall p\in [1,\infty]\qquad  \|a\|_p=\sup_{\|x\|_q=1}|\braket{x|a}|\ .
\end{eqnarray}
%This implies that the space $L^q$ is dual to $L^p$ with respect to the $L^2$ inner product. 
 Similarly, from the reverse H\"{o}lder inequality in (\ref{reverseHolder}) we have
 \begin{eqnarray}\label{infnorm}
     \forall p\in (0,1)\qquad \|a\|_p=\inf_{\|x\|_q=1}|\braket{x|a}|\ .
 \end{eqnarray}
% Since every operator has a polar decomposition in terms of a positive operator and a partial isometry we have
% \begin{eqnarray}\label{LpArakiMas}
%     &&\forall p\in [1,\infty]\qquad \|\ket{a}\|_p=\sup_{\|x\|_q=1}|\braket{x|a}|\ .
%      \nn\\
%       &&\forall p\in (0,1)\qquad \|\ket{a}\|_p=\inf_{\|x\|_q=1}|\braket{x|a}|\ .
% \end{eqnarray}
The equations above generalize (\ref{dualL2}) and (\ref{dualL1}) to arbitrary $p$. The duality between $L^1$ and $L^\infty$ is a special case of the duality between $L^p$ and $L^q$. That is why the parameter $q$ is called the H\"{o}lder dual of $p$. 

The vector $\ket{a}$ is a purification of the unnormalized density matrix $aa^\dagger=a_+^2$ of the algebra:
\begin{eqnarray}
   \braket{a|b|a}=\tr(b aa^\dagger)\ .
\end{eqnarray}
All vectors $\ket{a_+ u}$ purify the same state $a_+^2$. To make the purification unique, we define an anti-linear swap map $J_e$ in the basis of $\ket{k}$ in the definition of the vector $\ket{e}$:
\begin{eqnarray}
   J_e\ket{k,k'}=\ket{k',k}\ .
\end{eqnarray}
The map $\mathcal{J}_e(a)=J_e a J_e$ is an anti-unitary from $\mA$ to the commutant algebra $\mA'$ that acts as
\begin{eqnarray}\label{modularconjug}
   J_e(a\otimes \mI)J_e=(\mI\otimes (a^\dagger)^T)
\end{eqnarray}
and the transpose matrix $a^T$ defined in the $\{\ket{k}\}$ basis satisfies the equation 
\begin{eqnarray}
   (a\otimes \mI)\ket{e}=(\mI\otimes a^T)\ket{e}\ .
\end{eqnarray}
% The vectors $\ket{a_+}$ are the eigenvectors of $J_e$:
% \begin{eqnarray}
%   J_e(a_+\otimes \mI)\ket{e}=(\mI\otimes a_+)J_e\ket{e}=(a_+\otimes \mI)\ket{e}\ .
% \end{eqnarray}
The only purification of the unnormalized density matrix $aa^\dagger$ that is invariant under $J_e$ is  
\begin{eqnarray}\label{positiveJ}
\ket{a_+}=(a_+^{1/2}\otimes (a_+^{1/2})^T)\ket{e}\ .
\end{eqnarray}
The set of such vectors is called the natural cone in $\mH_e$ that we denote by $P_e^\natural$. Vectors in the natural cone are in one-to-one correspondence with the unnormalized density matrices $aa^\dagger=a_+^2$. 

To understand the $L^p$ spaces better we define the relative modular operators corresponding to algebra $\mA$:
%and $\mA'$:
\begin{eqnarray}
   &&\Delta_{\psi|\omega}\equiv \psi\otimes \omega^{-1}\ .
%   \nn\\
%   &&\Delta_{|\psi;\mA'}\equiv \psi^{-1}\otimes =\Delta_{\psi|\rho;\mA}^{-1}\ .
\end{eqnarray}
%\NL{(do we need to define it for $\mA'$?)}
The vector $\ket{e}$ reduced to the algebras $\mA$ and $\mA'$ gives the identity operator as an unnormalized state. We use the notation
$\Delta_{\omega|e}=\omega\otimes \mI$.
The superoperator on $\mA$ that correspond to the relative modular operator is
\begin{eqnarray}
   %&&\Delta_{\omega|e}^\theta a\ket{e}=\mathcal{D}^\theta_{\omega|e}(a)\ket{e}\nn\\
   &&\mathcal{D}_{\psi|\omega}(a)=\psi a\omega^{-1}\in \mA\ .
   %\nn\\ 
   %&&\mathcal{D}_{\omega|e}^{\theta}(a')=a' \omega^\theta\in \mA'\ .
\end{eqnarray}
% Every operator in the algebra has a right polar decomposition that we can think of as $v \omega^{1/2}$ with $\omega$ an unnormalized density matrix.
To every density matrix $\omega$ we can associate an operator $\ket{\omega}\bra{e}\in L^1$ with unit $1$-norm and a vector in $L^p$ 
\begin{eqnarray}
&& \ket{\omega^{1/p}}=\Delta_{\omega|e}^{1/p}\ket{e}= \Delta_{\omega|e}^{1/p-1/2}\ket{\omega^{1/2}}\ .
\end{eqnarray} 
with unit $p$-norm. We can think of the $L^p$ space as the space of vector $u\ket{\omega^{1/p}}$ for arbitrary $\omega$ and unitary $u$. 

We use the H\"{o}lder inequality to write the $p$-norm of a vector as
\begin{eqnarray}
   \|\ket{a}\|_{2p}^2=\|aa^\dagger\|_p&&=\sup_{\|\ket{\psi^{1/2}}\|_q=1}|\braket{\psi^{1/2}|aa^\dagger}|=\sup_{\|\ket{\omega^{1/2}}\|=1}|\braket{e|\Delta^{1/q}_{\omega|e}|aa^\dagger}|\nn\\
  &&=\sup_{\|\ket{\omega^{1/2}}\|=1}|\braket{a|\Delta_{\omega|e}^{1/q}|a}| = \sup_{\|\ket{\omega^{1/2}}\|=1}\|\Delta_{\omega|e}^{\frac{1}{2}-\frac{1}{2p}}\ket{a}\|^2\ .
   \end{eqnarray}
Above we have used the fact that any vector in the natural cone $\ket{\psi^{1/2}}\in L^q$ can be written as $\Delta_{\omega|e}^{1/q}\ket{e}$.\footnote{Since $\ket{aa^\dagger}$ is in the natural cone it follows from (\ref{Holdereq}) that the vector that saturates the H\"{o}lder inequality is also in the natural cone. Therefore, in the definition of the $q$-norm in (\ref{sup}) for $\ket{aa^\dagger}$ we can restrict the supremum to the vectors $\ket{\psi^{1/2}}$ in the natural cone.}
After a change of variables from $2p\to p$ we have
\begin{eqnarray}\label{Lpnormsrel}
    &&\|\ket{a}\|_p=\sup_{\|\ket{\omega^{1/2}}\|=1}\|\Delta_{\omega|e}^{\frac{1}{2}-\frac{1}{p}}\ket{a}\|\qquad \forall p\in [2,\infty]\ .
    %\nn\\
    % &&\|\ket{a}\|_p=\inf_{\ket{\psi}\in\mH_e}\|\Delta_{\psi|e}^{\frac{1}{2}-\frac{1}{p}}\ket{a}\|\qquad \forall p\in (0,1)\ .
\end{eqnarray}
We remind the reader that the norm of the vectors in the Hilbert space on the right-hand-side of the equations above is the $L^2$ norm.
Similarly, repeating (\ref{Lpnormsrel}) for the range $p\in (0,1)$ using (\ref{infnorm}) we obtain
  \begin{eqnarray}
  &&\|\ket{a}\|_p=\inf_{\|\ket{\omega^{1/2}}\|=1}\|\Delta_{\omega|e}^{\frac{1}{2}-\frac{1}{p}}\ket{a}\|\qquad \forall p\in [1,2)\ .     
  \end{eqnarray}

%\NL{(Was the paragraph above necessary? Do we use the natural cone?)}

The R\'enyi entropy of a normalized density matrix $\omega$ on $\mA$ can be written in terms of the $2p$-norm of the vector $\ket{\omega^{1/2}}\in \mH_e$:
\begin{eqnarray}\label{Renyip}
    S_p(\omega)\equiv \frac{2p}{1-p}\log\|\ket{\omega^{1/2}}\|_{2p}= \frac{p}{1-p}\log\|\omega\|_p =\frac{1}{1-p}\log\tr(\omega^p)\ .
\end{eqnarray}
Since $p$-norms of the vector $u'\ket{\omega^{1/2}}$ is independent of the unitary $u'$ the definition above defines the R\'enyi entropy for the reduced state $\omega$ on $\mA$ for any vector $\ket{\Omega}\in \mH_e$ %to define the R\'enyi entropy of its reduced state $\psi$ on $A$:
\begin{eqnarray}
   S_p(\omega)=\frac{2p}{1-p}\log \|\ket{\Omega}\|_{2p}\ .
\end{eqnarray}
 The normalized vector $d^{-1/2}\ket{e}$ corresponds to the maximally mixed density matrix and maximizes the R\'enyi entropy. In the limit $p\to 1$, we obtain the von Neumann entropy:
 \begin{eqnarray}
    S(\omega)=-2\lim_{p\to 1}\p_p\|\ket{\omega^{1/2}}\|_{2p}\ .
 \end{eqnarray}

\section{Operator \texorpdfstring{$L^p_\omega$}{} spaces}\label{sec:Lprho}

The construction of the $L^p$ spaces in the last section used the unnormalized vector $\ket{e}$. In an infinite dimensional algebra, this vector is not normalizable. The first step in generalizing the discussion of the last section to infinite dimensions is to replace the maximally mixed state with an arbitrary density matrix $\omega$:
\begin{eqnarray}
 \omega=\sum_k p_k\ket{k}\bra{k}
\end{eqnarray}
that for simplicity we will assume to be full rank. Not every infinite-dimensional algebra admits density matrices, however as we discuss in section \ref{sec:infinite}, the construction presented in this section generalizes to the algebras with no density matrices. We remind the reader that in our notation $\omega$ and $\psi$ are the reduced density matrices on $\mA$ corresponding to the vectors $\ket{\Omega}$ and $\ket{\Psi}$ in the Hilbert space $\mH_e=\mH_A\otimes \mH_{A'}$.

We generalize the definition of $p$-norm in (\ref{Lpnorm}) to a $(p,\omega)$-norm:
\begin{eqnarray}\label{Lprhonorm}
&&\|a\|_{p,\omega}\equiv \|a\ket{\omega^{1/p}}\|_p=\|a\omega^{1/p}\|_p\ .
%= \|(\Delta'_{\omega|e})^{1/p}\ket{a}\|_p\ . 
\end{eqnarray}
Note that the $(p,\omega)$-norm is no longer invariant under $a\to a^\dagger$.\footnote{We can define an alternate $(p,\omega,*)$-norm to be
\begin{eqnarray}\label{alternatenorm}
\|a\|_{p,\omega,*}\equiv\|a^\dagger\|_{p,\omega}=\|\omega^{1/p}a\|_p=\|\Delta_{\omega|e}^{1/p}\ket{a}\|_p\ .
\end{eqnarray}
As opposed to the $p$-norm the $(p,\omega)$-norm is not invariant under $a\to u a v$ with $u$ and $v$ unitaries. Instead, we have 
\begin{eqnarray}
   &&\|ua\|_{p,\omega}=\|a\|_{p,\omega},\qquad \|au\|_{p,\omega,*}=\|a\|_{p,\omega,*}\ .
\end{eqnarray}
In other words, for unitaries $u\in\mA$ and $u'\in \mA'$ we have 
\begin{eqnarray}
 &&\|u\ket{\Psi}\|_{p,\omega}=\|\ket{\Psi}\|_{p,\omega},\qquad 
\|u'\ket{\Psi}\|_{p,\omega,*}=\|\ket{\Psi}\|_{p,\omega,*}\ .
\end{eqnarray}
More generally, one can define  the Kosaki $(p,\sigma,\omega)$-norms 
 $\|a\|_{p,\sigma,\omega}=\|\sigma^{1-1/p}a \omega^{1/p}\|_p$ \cite{kosaki1984applications}.}
 Consider the $*$-representation $\pi(a)=a\otimes 1_R$ with some auxiliary system $R$. The $(p,\omega)$-norm satisfies the equality
 \begin{eqnarray}
  \|(a\otimes 1_R)\|_{p,\omega_{AR}}=\|a\|_{p,\omega}
 \end{eqnarray}
 if $\omega_{AR}=\omega_A\otimes \sigma_R$.

We consider the representation map 
\begin{eqnarray}
   &&a\to \ket{a}_\omega\equiv a\ket{\omega^{1/2}}\nn\\
   &&\ket{\omega^{1/2}}=\sum_k \sqrt{p_k}\ket{k,k}\ .
\end{eqnarray}
Since $\omega$ is full rank this representation is faithful.
We call $\mH_\omega$ the GNS Hilbert space and sometimes refer to it as the $L^2_\omega$ Banach space because the $L^2_\omega$ norm is the Hilbert space norm: 
\begin{eqnarray}\label{2omega}
\|a\ket{\omega^{1/2}}\|_{2,\omega}=\|a\ket{\omega^{1/2}}\|_2\ .
\end{eqnarray}
Since the $(\infty,\omega)$-norm is the same as the $\infty$-norm the algebra itself is the $L^\infty_\omega$ space. The $L^1_\omega$ space is the space of operators $\omega_a=a\ket{\omega}\bra{e}$ with the $L^1$ norm.
 Note that as opposed to the $p$-norm, for the $(p,\omega)$-norms we have the hierarchy $L^\infty_\omega\subseteq L^2_\omega\subseteq L^1_\omega$ when $\omega$ is a normalized density matrix because of the inequalities
 \begin{eqnarray}
    &&\|a\|_1\geq \|a\|_2\geq \|a\|_\infty\nn\\
    &&\|a\|_{1,\omega}\leq \|a\|_{2,\omega}\leq \|a\|_{\infty,\omega}\ .
    \end{eqnarray}

The vector $\ket{a}_\omega$ in the Hilbert space $\mH_\omega$ corresponds to the state (density martix of $\mA$) $\omega_a=a\omega a^\dagger$. However, given a density matrix 
there are many vectors in $\mH_\omega$ that purify it. In the last section, we used the modular conjugation operator $J_e$ to fix a canonical vector for each density matrix. 
To fix a canonical vector we start with the map
\begin{eqnarray}
   \psi\to \ket{\psi^{1/2}}=(\psi^{1/2}\omega^{-1/2})\ket{\omega^{1/2}}\in \mH_\omega\ .
\end{eqnarray}
Any state of the form 
\begin{eqnarray}
  (\psi^{1/2}u \omega^{-1/2})\ket{\omega^{1/2}}
\end{eqnarray}
for unitary $u$ has the same density matrix $\psi$. 
% The relative modular operator $\Delta_{\psi|\omega}=\psi\otimes \omega^{-1}$ corresponds to the  superoperator $\mathcal{D}_{\psi|\omega}(a)=\psi a \omega^{-1}$ and
% \begin{eqnarray}
%   &&\ket{\mathcal{D}_{\psi|\omega}(u)}_\omega=\Delta_{\psi|\omega}^{1/2}(u\otimes \mI)\ket{\omega^{1/2}}\ .
% \end{eqnarray}
%for unitary $u$ has the same reduced state $\psi_a$.
To make the correspondence between the density matrices and their purification one-to-one we introduce the modular conjugation operator $J_\omega$ that acts as (\ref{modularconjug}) in the eigenbasis of $\omega$.
From the argument in (\ref{positiveJ}) it is clear that the vector $\ket{\psi^{1/2}}$ is the only $J_\omega$ invariant vector representative of the density matrix $\psi$.
Therefore, there is a one-to-one correspondence between the density matrices $\psi$ and the vectors 
\begin{eqnarray}
   \ket{\psi^{1/2}}=\Delta_{\psi|\omega}^{1/2}\ket{\omega^{1/2}}
\end{eqnarray}
that are invariant under $J_\omega$. These vectors form the so-called natural cone $P^\natural_\omega$.
%\NL{(do we use this concept?)}

We define the $(p,\omega)$-norm of the vectors in the GNS Hilbert space $\mH_\omega$ to be
\begin{eqnarray}\label{Lpstar}
   &&\|a\ket{\omega^{1/2}}\|_{p,\omega}\equiv\|a\|_{p,\omega}
   %\nn\\
\end{eqnarray}
so that the $(2,\omega)$-norm is the Hilbert space norm of $a\ket{\omega^{1/2}}$.
Note that $\ket{\omega^{1/2}}$ has unit $(p,\omega)$-norm for all $p$.\footnote{
We can also define the alternate $(p,\omega,*)$-norm of a vector
\begin{eqnarray}
   &&\|a\ket{\omega^{1/2}}\|_{p,\omega,*}\equiv\|a\|_{p,\omega,*}\equiv\|a^\dagger\ket{\omega^{1/2}}\|_{p,\omega}\ .
\end{eqnarray}
The $(2,\omega,*)$ is the Hilbert space norm of $a^\dagger\ket{\omega^{1/2}}$.
The $(p,\omega,*)$-norm of a vector has the advantage that it is independent of unitary rotations $u'\in \mA'$:
\begin{eqnarray}
   \|u'a\ket{\omega^{1/2}}\|_{p,\omega,*}\equiv \|u'a\|_{p,\omega,*}=\|a\ket{\omega^{1/2}}\|_{p,\omega,*}\ .
\end{eqnarray}
Therefore, it only depends on the reduced state on $A$ that is $aa^\dagger$, and not a particular purification choice $u'\ket{a}$.}

To every density matrix $\psi$ we can canonically associate a unique operator $\ket{\psi \omega^{-1/2}}\bra{\omega^{1/2}}\in L^1_\omega$ with unit $1$-norm and a unique vector in $L^p_\omega$ with unit $(p,\omega)$-norm:
\begin{eqnarray}
   \psi\to  \ket{\psi^{1/p}\omega^{1/2-1/p}}=\Delta_{\psi|\omega}^{1/p}\ket{\omega^{1/2}}=\Delta_{\psi|\omega}^{1/p-1/2}\ket{\psi^{1/2}}\ .
\end{eqnarray}
As we vary from $p=2$ to $p=\infty$ the vector above interpolates between $\ket{\psi^{1/2}}$ with unit $(2,\omega)$-norm and $\ket{\omega^{1/2}}$ with unit $(\infty,\psi)$-norm.
Note that if $\psi$ is not normalized we have
\begin{eqnarray}\label{omegap}
 \|\Delta_{\psi|\omega}^{1/p}\ket{\omega^{1/2}}\|_{p,\omega}=\|\psi\|_1^{1/p}
\end{eqnarray}
which is independent of $\omega$. 
Since $\omega$ is invertible and $L^r_\omega\subseteq L^p_\omega$ for any $p\leq r$ the vector
\begin{eqnarray}
\Delta_{\psi|\omega}^{\theta/p}\ket{\omega^{1/2}}\in L^p_\omega
\end{eqnarray}
for any $\theta\in [0,1]$.\footnote{Note that in finite dimensions we can take $\theta>1$ as well. However, in this work, we restrict to the range because it generalizes to infinite dimensions.}
In fact, we can extend $\theta$ to the complex plane $z=\theta+it$ because 
\begin{eqnarray}
 &&\Delta_{\psi|\omega}^{it}\ket{\omega^{1/2}}=(D\psi:D\omega)_t\ket{\omega^{1/2}}\nn\\
 &&(D\psi:D\omega)_t\equiv \Delta_{\psi|\omega}^{it}\Delta_\omega^{-it}
\end{eqnarray}
and the cocycle $(D\psi:D\omega)_t$ is a partial isometry in the algebra for all real values of $t$. When $\psi$ is full rank the cocycle is a unitary operator.

As we saw in the last section, the H\"{o}lder inequality helps bound the $p$-norm in terms of simpler norms such as the $2$-norm and $\infty$-norm. In section (\ref{sec:multiineq}), we will prove the following H\"{o}lder inequality for the $(p,\omega)$-norms
\begin{eqnarray}
 &&\|\Delta_{\psi_0|\omega}^{1/p_0}\Delta_{\psi_1|\omega}^{1/p_1}\ket{\omega^{1/2}}\|_{r,\omega}\leq \|\psi_0\|_1^{1/p_0}\|\psi_1\|_1^{1/p_1}\nn\\
 &&\frac{1}{p_0}+\frac{1}{p_1}=\frac{1}{r}\ .
\end{eqnarray}
% We will prove this as a $\theta=1/2$ case of the inequality 
% \begin{eqnarray}\label{holdervN}
%   &&\|\Delta_{\psi_0|\omega}^{(1-\theta)/p_0}\Delta_{\psi_1|\omega}^{\theta/p_1}\ket{\omega^{1/2}}\|_{p_\theta,\omega}\leq \|\Delta_{\psi_0|\omega}^{1/p_0}\ket{\omega^{1/2}}\|^{1-\theta}_{p_0,\omega}\|\Delta_{\psi_1|\omega}^{1/p_1}\ket{\omega^{1/2}}\|_{p_1,\omega}^\theta\nn\\
%   &&\frac{1}{p_\theta}=\frac{1-\theta}{p_0}+\frac{\theta}{p_1}
% \end{eqnarray}
% for $\theta\in (0,1)$. \NL{Need to be checked. $\theta$ needs to be changed?} The right-hand-side of the equation above is equal to one if the states are normalized.

\begin{figure}[t!]
 \centering
 \includegraphics[scale=0.38]{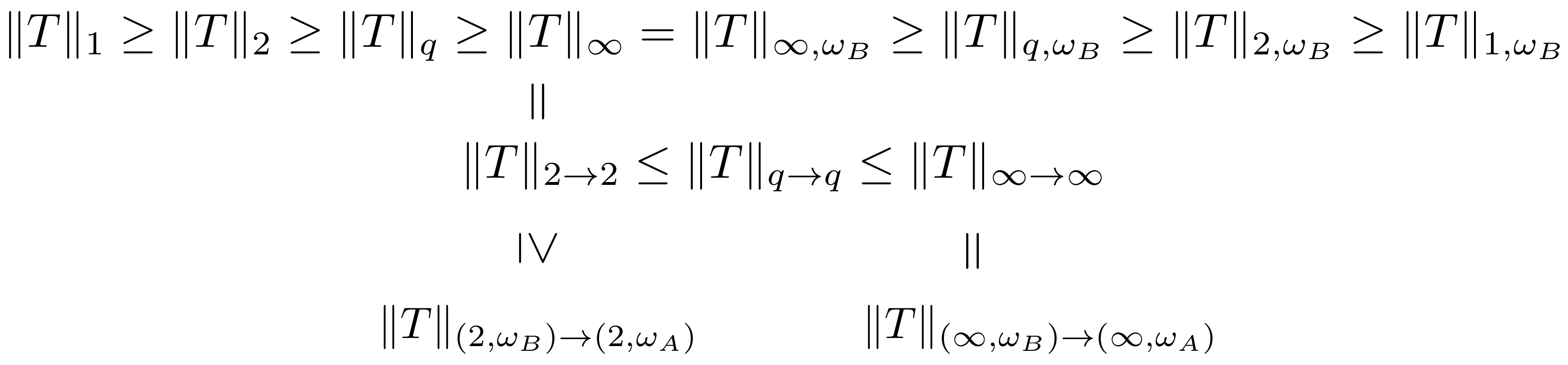}
 \caption{\small{The figure shows the hierarchy of norms for a linear operator $T:\mH_B \to \mH_A$} and $2<q<\infty$. The inequality between $\|T\|_{2\to 2}$ and $\|T\|_{(2,\omega_B)\to (2,\omega_A)}$ saturates when the size of $\mH_B$ and $\mH_A$ are the same.}
 \label{fig:hierarchy}
\end{figure}

Similarly, it is often helpful to relate the $(p,\omega)\to (p,\omega)$ norms of superoperators in (\ref{supernorm}), or equivalently those of their corresponding operators in the GNS Hilbert space in (\ref{supernorm2}). This is achieved using an inequality established by the Riesz-Thorin interpolation theorem that we prove in appendix \ref{App:Riezs}. The theorem says that for $2\leq p_0,p_1$ and $\theta\in [0,1]$ and any operator $T:\mH_A\to \mH_B$ we have 
\begin{eqnarray}\label{Riezs}
   &&\|T\|_{(p_\theta,A)\to (p_\theta,B)}\leq \|T\|_{(p_0,A)\to (p_0,B)}^{1-\theta}\|T\|_{(p_1,A)\to (p_1,B)}^\theta\nn\\
   &&\frac{1}{p_\theta}=\frac{1-\theta}{p_0}+\frac{\theta}{p_1}\ .
\end{eqnarray}
Consider a contraction $F:\mH_A\to \mH_B$\footnote{We remind the reader that a contraction is defined with respect to the infinity norm, and not any other norms we discuss in this work.} where $\mH_A\equiv \mH_{\omega_A}$ and $\mH_B\equiv \mH_{\omega_B}$ are the GNS Hilbert spaces of states $\omega_A$ and $\omega_B$, respectively. Since $(2,A)$-norm is the Hilbert space norm and $(\infty,A)$ norm is the $\infty$-norm, by the same argument as in (\ref{contractinf}), we have
\begin{eqnarray}\label{contract}
   &&\|F\ket{a}_{\omega_A}\|_{2,B}\leq \|\ket{a}_{\omega_A}\|_{2,A}\nn\\
   &&\|F\ket{a}_{\omega_A}\|_{\infty,B}\leq \|\ket{a}_{\omega_A}\|_{\infty,A}\ .
\end{eqnarray}
Then, the Riesz-Thorin inequality in (\ref{Riezs}) implies that contractions cannot increase the $(p,\omega)$-norm of a vector for $p\geq2$, i.e.
\begin{eqnarray}\label{contractRiezs}
 \|F\|_{(p,A)\to (p,B)}\leq 1\ .
\end{eqnarray}
The above result plays a central role in our proof of the data processing inequality. See figure \ref{fig:hierarchy} for the relation between different norms.

In (\ref{Lpnormsrel}) we used the H\"{o}lder inequality to rewrite the $p$-norm of the vectors as a variational expression in the Hilbert space.
In constructing the GNS Hilbert space we replaced $\ket{e}$ with the state $\ket{\omega^{1/2}}$ and defined the vectors $\ket{a}_\omega=a\ket{\omega^{1/2}}$. The definition of the $L^p_\omega$ norms in (\ref{Lpnormsrel}) generalizes to the GNS Hilbert space:
\begin{eqnarray}
    \|\ket{a}_\omega\|_{2p,\omega}^2=\|a\omega^{\frac{1}{2p}}\|_{2p}^2=\|a\omega^{1/p}a^\dagger\|_p&&=\sup_{\|\ket{\psi^{1/2}}\|=1}\|\Delta_{\psi|e}^{\frac{1}{2}-\frac{1}{2p}}\ket{a\omega^{1/(2p)}}\|^2\nn\\ &&=\sup_{\|\ket{\psi^{1/2}}\|=1}\|\Delta_{\psi|\omega}^{\frac{1}{2}-\frac{1}{2p}}\ket{a}_\omega\|^2\ .
\end{eqnarray}
After a change of variables from $2p\to p$ we find
\begin{eqnarray}\label{arakimasuda}
&&\|\ket{a}_\omega\|_{p,\omega}=\sup_{\ket{\psi^{1/2}}\in\mH_\omega}\|\Delta_{\psi|\omega}^{\frac{1}{2}-\frac{1}{p}}\ket{a}_\omega\|\qquad \forall p\in [2,\infty]\nn\\
&&\|\ket{a}_\omega\|_{p,\omega}=\inf_{\ket{\psi^{1/2}}\in\mH_\omega}\|\Delta_{\psi|\omega}^{\frac{1}{2}-\frac{1}{p}}\ket{a}_\omega\|\qquad \forall p\in [1,2)\ .
\end{eqnarray}
where $\ket{\psi^{1/2}}$ has unit norm. 
In \cite{araki1982positive}, Araki and Masuda observed that the definition of the $(p,\omega)$-norm above generalize to any von Neumann algebra, even to those that do not admit a trace such as the local algebra of QFT. We will come back to this in section \ref{sec:infinite}.

 \subsection{Two-state R\'enyi divergences}
 
Now, we are ready to define the distinguishability measures using the $L^p_\omega$ norm of the vectors in the GNS Hilbert space. 
We define the Petz divergences in terms of the Hilbert space norm of the interpolating vector
\begin{eqnarray}\label{Petzdiv}
   D_{1/p}(\psi\|\omega)=\frac{2p}{1-p}\log\|\Delta_{\psi|\omega}^{1/(2p)}\ket{\omega^{1/2}}\|_{2,\omega}
\end{eqnarray}
and the sandwiched R\'enyi divergences using the $(p,\omega)$-norm of the vector $\ket{\psi^{1/2}}$
\cite{wilde2014strong,muller2013quantum}
\begin{eqnarray}\label{Renyiprho}
    S_p(\psi\|\omega)\equiv \frac{2p}{p-1}\log\|\ket{\psi^{1/2}}\|_{2p,\omega}=\frac{p}{p-1}\log\|\omega^{-\frac{1}{2q}}\psi\omega^{-\frac{1}{2q}}\|_p
\end{eqnarray}
for $p\in [1/2,\infty]$.\footnote{Cases $p=1$ and $p=\infty$ are defined as limits $p\to 1$ and $p\to\infty$.} These divergences are the generalizations of the R\'enyi entropy in (\ref{Renyip}) to the GNS Hilbert space. Their asymmetry has to do with the fact that the reference state $\omega$ is used to construct the GNS Hilbert space. These two-state R\'enyi divergences satisfy the data processing inequality \cite{beigi2013sandwiched,frank2013monotonicity,lashkari2019constraining}.
The $p\to 1$ limit of both quantities gives the relative entropy \cite{wilde2018optimized}
\begin{eqnarray}
   S(\psi\|\omega)=-2\lim_{p\to 1}\p_p\|\ket{\psi^{1/2}}\|_{2p,\omega}\ .
   %=-2\lim_{p\to 1}\p_p\|\Delta_{\psi|\omega}^{1/(2p)}\ket{\omega^{1/2}}\|\ .
\end{eqnarray}
Since we will be always working in the GNS Hilbert space $\mH_\omega$ we simplify our notation by introducing $\ket{\Omega}\equiv \ket{\omega^{1/2}}$. 
The vector $\ket{\psi^{1/2}}$ is a purification of $\psi$ which is symmetric under $J_\Omega$. It can be written as
\begin{eqnarray}
 \ket{\psi^{1/2}}=\Delta_{\psi|\omega}^{1/2}\ket{\Omega}\ .
\end{eqnarray}
The definitions in (\ref{Petzdiv}) and (\ref{Renyiprho}) are independent of the purification of $\psi$.
To see this, we first define the relative modular operator for an arbitrary vector $\ket{\Psi}$ 
 \begin{eqnarray}
  \Delta_{\Psi|\Omega}\equiv\psi_A\otimes \omega_{A'}^{-1}
 \end{eqnarray}
so that it remains unchanged for other purifications of $\psi$:
\begin{eqnarray}\label{relmodeunitary}
   \Delta_{u'\Psi|\Omega}=\Delta_{\Psi|\Omega}\ .
\end{eqnarray}
For an arbitrary vector $\ket{\Psi}\in \mH_\omega$ we can write the divergences in (\ref{Petzandsandwiched}) as
\begin{eqnarray}\label{Renyivector}
   &&D_{1/p}(\psi_A\|\omega_A)\equiv \frac{2p}{1-p}\log\|\Delta_{\Psi|\Omega}^{1/(2p)}\ket{\Omega}\|_{2,\Omega}\nn\\
   &&S_p(\psi_A\|\omega_A)\equiv \frac{2p}{p-1}\log\|\Delta_{\Psi|\Omega}^{1/2}\ket{\Omega}\|_{2p,\Omega}\ .
\end{eqnarray}
We also define the $(p,\Omega)$-norm in the GNS Hilbert space $\mH_\Omega$ using
\begin{eqnarray}\label{arakimasuda2}
   &&\|\ket{\Psi}\|_{p,\Omega}=\sup_{\|\ket{\chi}\|=1}\|\Delta_{\chi|\Omega}^{1/2-1/p}\ket{\Psi}\|\qquad p\in [2,\infty]\nn\\
   &&\|\ket{\Psi}\|_{p,\Omega}=\inf_{\|\ket{\chi}\|=1}\|\Delta_{\chi|\Omega}^{1/2-1/p}\ket{\Psi}\|,\qquad p\in [1,2)\ .
\end{eqnarray}
To interpolate between the two divergences following \cite{audenaert2015alpha} we introduce the $(\theta,r)$-entropies
\begin{eqnarray}\label{alphaz}
   S_{\theta,r}(\psi_A\|\omega_A)=\frac{-2r}{1-\theta}\log\|\Delta_{\Psi|\Omega}^{\theta/(2r)}\ket{\Omega}\|_{2r,\Omega}
\end{eqnarray}
for the range $r\in [1/2,\infty]$ and $\theta\in [0,1]$. Even though in matrix algebras one can extend beyond this range we limit our discussion to  this limited range because outside of this range, in infinite dimensions, the $(\theta,r)$-R\'enyi divergences might not be finite. We postpone a study of the extended range to future work. 

In matrix algebras, the expression in (\ref{alphaz}) becomes
\begin{eqnarray}
S_{\theta,r}(\psi_{A}\|\omega_{A})
&&=\frac{2r}{\theta-1}\log\|\psi_A^{\theta/(2r)}\omega_A^{(1-\theta)/(2r)}\|_{2r}  \nn \\ && =\frac{1}{\theta-1}\log\tr\left[ \lb \omega_A^{\frac{1-\theta}{2r}}\psi_A^{\frac{\theta}{r}}\omega_A^{\frac{1-\theta}{2r}}\rb^r\right]
\end{eqnarray}
where in the first equality we have used
\begin{eqnarray}
   (1\otimes \omega_{A'})\ket{\Omega}=(\omega_A\otimes 1)\ket{\Omega}\ .
\end{eqnarray}
It follows from the definition in (\ref{alphaz}) that the $(\theta,\theta)$-R\'enyi divergences is the $\theta$-sandwiched R\'enyi divergence and the $(\theta,1)$-R\'enyi divergences is the $\theta$-Petz divergence. In the remainder of this work, we suppress the subscript $A$ unless there is potential for confusion. Note that the matrix algebra expression enjoys the symmetry
\begin{eqnarray}\label{Srsym}
 (1-\theta) S_{\theta,r}(\psi\|\omega)=\theta S_{1-\theta, r}(\omega\|\psi)\ .
\end{eqnarray}
In the limit $r\to \infty$ we can use the Lie-Trotter formula 
% \cite{}
 \begin{eqnarray}
  \lim_{r\to \infty}\lb e^{a_1/r} e^{a_2/r}\rb^r=e^{a_1+a_2}
\end{eqnarray}
for self-adjoint operators $a_1,a_2$ to write
\begin{eqnarray}\label{rinf}
 \lim_{r\to \infty} S_{\theta,r}(\psi\|\omega)&&=\frac{1}{1-\theta}\log\tr\lb e^{\theta \log\psi+(1-\theta)\log\omega}\rb\ .
%  \nn\\
%  &&=\frac{1}{1-\theta}\log\braket{\Psi|e^{(1-\theta)\log\Delta_\Psi+\theta \Delta_{\psi|\Psi}}|\Psi}\ .
\end{eqnarray}
A larger class of two-state R\'enyi divergences one can consider is
\begin{eqnarray}\label{Sfr}
 S^f_r(\psi\|\omega)=-2r\log\|f(\Delta^{1/r}_{\Psi|\Omega})^{1/2}\ket{\Omega}\|_{2r,\Omega}
\end{eqnarray}
where $f$ is an operator monotone function.\footnote{A function $f:(0,\infty)\to \mathbb{R}$ is called operator monotone if for positive operators $X$ and $Y$ the inequality $X\leq Y$ implies $f(X)\leq f(Y)$.} In the next subsection, we show that these measures satisfy the data processing inequality. They are related to the $f$-divergences and the Petz quasi-entropies \cite{petz1985quasi,petz1986quasi,wilde2018optimized,lashkari2019constraining}. A few examples of the operator monotone functions are
\begin{enumerate}
    \item $f(x)=x^\alpha$ with $\alpha\in (0,1)$. 
    \item $f(x)=-x^{-\alpha}$ with $\alpha\in (0,1)$. 
    \item $f(x)=x\log x$
    \item $f(x)=\log x$
\end{enumerate}
For concreteness and the fact that at times we need $f(1)=1$, we will be mostly concerned with the first case: $f(x)=x^\alpha$. However, we prove the data processing inequality for a general operator monotone function $f$.

\subsection{Data processing inequality for $(\theta,r)$-R\'enyi divergences}\label{sec:dataprocess}

Consider  a quantum channel $\Phi^*$ that sends the density matrices $\psi_A$ and $\omega_A$ to $\psi_B=\Phi^*(\psi_A)$ and $\omega_B=\Phi^*(\omega_A)$, respectively. We consider the GNS Hilbert spaces corresponding to $\omega_A$ and $\omega_B$ and call them $\mH_A$ and $\mH_B$. We have
\begin{eqnarray}\label{defomegaB}
 \braket{\Omega_B|b|\Omega_B}=\braket{\Omega_A|\Phi(b)|\Omega_A}\ .
\end{eqnarray}

In this subsection, we prove the data processing inequality for the $(\theta,r)$-R\'enyi divergences in (\ref{alphaz}) and the divergences in (\ref{Sfr}) at $r\geq 1$:
\begin{eqnarray}\label{dataprocess2states}
&&S^f_r(\Phi^*(\psi_A)\|\Phi^*(\omega_A))\leq S^f_r(\psi_A\|\omega_A)\nn\\
&&   S_{\theta,r}(\Phi^*(\psi_A)\|\Phi^*(\omega_A))\leq S_{\theta,r}(\psi_A\|\omega_A)\ .
\end{eqnarray}
%This inequality is sometimes called the data processing inequality. 
In the range we are interested, the monotonicity of the $(\theta,r)$-R\'enyi divergences was first proved by \cite{hiai2013concavity}.\footnote{See theorem 2.1 of \cite{zhang2020wigner} for a proof of the data processing inequality in extended range of $(\theta,r)$ for matrix algebras.} 
In the Heisenberg picture, the quantum channel $\Phi^*$ is described by a unital CP map $\Phi:\mB\to B(\mH_A)$ that acts on the algebra. 
% We remind the reader that going from the Schr\"{o}dinger to the Heisenberg picture the domain and the range of the map are swapped. 
Note that the range of a CP map need not be the full algebra $B(\mH_A)$. For simplicity, sometimes we use the notation $\mA$ for the operators in $B(\mH_A)$.\footnote{In general, the range of a CP map is a $*$-closed subspace of observables inside $B(\mH_A)$, otherwise known as an operator system.} 

Consider a unital CP map $\Phi:\mB\to \mA$ and its corresponding contraction operator $F$ in the GNS Hilbert space:
%\footnote{See \cite{Furuya:2020tzv} for a review of the proof that $F$ is a contraction. We remind the reader that to compare with the discussion of contractions in section \ref{sec:Lprho} one has to switch $\mA$ and $\mB$ because in this section the CP map is from $\mB$ to $\mA$.}
\begin{eqnarray}
 \Phi(b)\ket{\Omega_A}=Fb\ket{\Omega_B}\ .
\end{eqnarray}
The monotonicity of the relative modular operator is the operator inequality\footnote{See  \cite{petz2003monotonicity}, and \cite{witten2018aps} for a review of its proof using the Tomita-Takesaki modular theory}:
\begin{eqnarray}\label{relativeMono}
   &&F^\dagger\Delta_{\Psi_A|\Omega_A} F\leq \Delta_{\Psi_B|\Omega_B}\ .
%   \nn\\
%   &&F^\dagger\Delta_{\omega_1|\rho;A}^\theta F\leq \Delta_{\omega_1|\rho;B}^\theta\ .
\end{eqnarray}
Choosing the function $f(x)=-(t+x)^{-1}$ that is operator monotone and operator convex\footnote{A function is called operator convex if $f(\theta X+(1-\theta) Y)\leq \theta f(X)+(1-\theta) f(Y)$.}  we obtain \cite{petz2003monotonicity}
\begin{eqnarray}
   F^\dagger\frac{1}{t+\Delta_{\Psi_A|\Omega_A}}F\geq \frac{1}{t+\Delta_{\Psi_B|\Omega_B}} \ .
\end{eqnarray}
Any operator monotone function $f$ can be expanded as \cite{schilling2012bernstein,bhatia2013matrix} 
\begin{eqnarray}\label{operatormonotone}
 f(X)=a + b X+ \int_0^\infty d\mu(t)\lb \frac{t}{t^2 + 1} - \frac{1}{t + X}\rb
\end{eqnarray}
for $a\in \mathbb{R}, b\geq 0$ and $\mu(t)$ a positive measure that satisfies\footnote{When $f(0) := \lim_{t\to 0} f(t) > -\infty$, we can write $f$ as
\begin{eqnarray}
    f(X)=f(0) + bX + \int_0^{\infty} d\mu(t) \lb \frac{1}{t} - \frac{1}{t+X} \rb
\end{eqnarray}
where $\mu(t)$ satisfies $\int_0^\infty \frac{1}{t+t^2}d\mu(t)<\infty$.
}
\begin{eqnarray}
 \int_0^\infty \frac{1}{t^2 + 1} d\mu(t)<\infty\ .
\end{eqnarray}
%\begin{eqnarray}\label{operatormonotone}
% f(X)=aX+b+\int_0^\infty dt \:\mu(t)\lb \frac{1}{t}-\frac{1}{t+X}\rb
%\end{eqnarray}
%for $a,b\geq 0$ and $\mu(t)$ a measure that satisfies 
%\begin{eqnarray}
% \int_0^1 \frac{dt}{t}\mu(t)+\int_1^\infty \frac{dt}{t^2}\mu(t)<\infty\ .
%\end{eqnarray}
Therefore, we have the inequality
\begin{eqnarray}\label{monoV}
 F^\dagger f(\Delta_{\Psi_A|\Omega_A})F\leq f(\Delta_{\Psi_B|\Omega_B})\ .
\end{eqnarray}
This implies
\begin{eqnarray}\label{FfFf}
f(\Delta_{\Psi_B|\Omega_B})^{-1/2} F^\dagger f(\Delta_{\Psi_A|\Omega_A}) F f(\Delta_{\Psi_B|\Omega_B})^{-1/2}\leq 1\ .
\end{eqnarray}
Define the operator 
\begin{eqnarray}\label{Ftheta}
&&F_f\equiv f(\Delta_{\Psi_A|\Omega_A})^{1/2} F f(\Delta_{\Psi_B|\Omega_B})^{-1/2}\ .
\end{eqnarray}
In appendix \ref{app:proof}, we show that (\ref{FfFf}) implies that $F_f$ is a contraction and satisfies
\begin{eqnarray}
&&\|F_f\|_{(p,\Omega_B)\to (p,\Omega_A)}\leq 1\ .
\end{eqnarray}
In the case of function $f(x)=x^\theta$ the integral representation in (\ref{operatormonotone}) is
\begin{eqnarray}
X^{\theta}=\frac{\sin(\pi \theta)}{\pi}\int_0^\infty dt\:t^\theta\lb \frac{1}{t}-\frac{1}{t+X}\rb\ .
\end{eqnarray}
which is equivalent to saying that $F_{\theta}$ satisfies:
\begin{eqnarray} \label{Ftheta2}
 &&\|F_{\theta}\|_{(p,\Omega_B)\to (p,\Omega_A)}\leq 1\nn\\
 &&F_{\theta}\equiv \Delta^{\theta/2}_{\Psi_A|\Omega_A}F \Delta^{-\theta/2}_{\Psi_B|\Omega_B}\ .
 \end{eqnarray}
This argument is similar to \cite{faulkner2020approximate}. 

To prove the monotonicity under a contraction we use a proof similar to the one presented in \cite{beigi2013sandwiched}:
\begin{eqnarray}\label{dataprooftwostates}
 \|f(\Delta_{\Psi_A|\Omega_A})^{1/2}\ket{\Omega_A}\|_{2r,\Omega_A}&& =\|F_f f(\Delta_{\Psi_B|\Omega_B})^{1/2}\ket{\Omega_B}\|_{2r,\Omega_A}\nn\\
%  &&=\sup_{\|\ket{\chi}\|_{(2s,B)}=1}|\braket{W_\theta^\dagger\chi|\Delta_{\psi|\omega;B}^\theta |\omega_B}|\nn\\
 &&\leq \|F_f\|_{(2r,\Omega_B)\to (2r,\Omega_A)}\|f(\Delta_{\Psi_B|\Omega_B})^{1/2}\ket{\Omega_B}\|_{2r,\Omega_B}\nn\\
 &&\leq \|f(\Delta_{\Psi_B|\Omega_B})^{1/2}\ket{\Omega_B}\|_{2r,\Omega_B}
\end{eqnarray}
where we have used the definition of the $(2r,\Omega_B)\to (2r,\Omega_A)$ norm for the contraction $F_f$ and the fact that it is less than one. 
We obtain the data processing inequalities in (\ref{dataprocess2states}) in the range $r\geq 1$:\footnote{We restrict to $r\geq 1$ as we proved the Riesz-Thorin theorem for this range in appendix \ref{App:Riezs}.}
\begin{eqnarray}\label{dataprocessing}
&&  S_{r}^f(\psi_B\|\omega_B)\leq S_{r}^f(\psi_A\|\omega_A)\ .
% &&  S_{\theta,r}(\psi_B\|\omega_B)\leq S_{\theta,r}(\psi_A\|\omega_A)\ .
 \end{eqnarray}
 In the case of $(\theta,r)$-R\'enyi divergences we find
 \begin{eqnarray}
  S_{\theta,r}(\psi_B\|\omega_B)\leq S_{\theta,r}(\psi_A\|\omega_A)
 \end{eqnarray}
 for $r\geq 1$ and $\theta\in [0,1]$. In appendix \ref{app:extended}, we
 show that if $\omega\leq c\psi$ for some constant $c$ the vector
 \begin{eqnarray}
  \Delta_{\Psi|\Omega}^{-\theta}\ket{\Omega}\in L^{2r}_\omega
 \end{eqnarray}
 in the extended range $\theta\in [-1,1]$ and $r\geq 1$.
 To prove the data processing inequality in (\ref{dataprooftwostates}) we used the contraction in (\ref{Ftheta2}):
\begin{eqnarray}
 F_\theta=\Delta_{\Psi_A|\Omega_A}^{\theta/2}F \Delta_{\Psi_B|\Omega_B}^{-\theta/2}\ .
\end{eqnarray}
The $\dagger$ of this operator is also a contraction
\begin{eqnarray}
 F_\theta^\dagger=\Delta_{\Psi_B|\Omega_B}^{-\theta/2}F^\dagger\Delta_{\Psi_A|\Omega_A}^{\theta/2}\ .
\end{eqnarray}
Therefore, we have 
\begin{eqnarray}
 \|\Delta_{\Psi|\Omega;B}^{-\theta/2}\ket{\Omega_B}\|_{2r,\Omega_B}=\|F^\dagger_\theta\Delta_{\Psi|\Omega;A}^{-\theta/2}\ket{\Omega_A}\|_{2r,\Omega_B}\leq \|\Delta_{\Psi|\Omega;A}^{-\theta/2}\ket{\Omega_A}\|_{2r,\Omega_A}
\end{eqnarray}
which says that the measure
\begin{eqnarray}
 S_{\theta,r}(\psi\|\omega)=\frac{-2r\:\text{sign}(\theta)}{1-\theta}\log \|\Delta^{\theta/(2r)}_{\Psi|\Omega}\ket{\Omega}\|_{2r,\Omega}
\end{eqnarray}
 satisfies the data processing inequality in the extended range $\theta\in (-1,1)$. Another way to define a measure with an extended range of monotonicity $\theta\in (-1,1)$ is
\begin{eqnarray}\label{negativetheta}
 \hat{S}_{\theta,r}(\psi\|\omega)\equiv \frac{-2r}{\theta(1-\theta)}\log \|\Delta^{\theta/(2r)}_{\Psi|\Omega}\ket{\Omega}\|_{2r,\Omega}\ .
\end{eqnarray}
Note that this measure no longer vanishes at $\theta\to 0$. For instance, when $r=1$ it corresponds to a modification of the Petz divergence
\begin{eqnarray}
 \frac{-2}{\theta(1-\theta)}\log\|\psi^\theta \omega^{1-\theta}\|
\end{eqnarray}
that interpolates between the relative entropy $S(\omega\|\psi)$ at $\theta\to 0$ and $S(\psi\|\omega)$ at $\theta\to 1$. The measures defined above satisfy the data processing inequality and vanishes for identical states, hence they are non-negative.\footnote{Consider the CP map that sends all states to the same $\omega_B$. After the channel the measure is zero. Since it has not increased, it was non-negative before applying the channel.}
 
In general, when $\theta>1$ we are not guaranteed that $\Delta_{\Psi|\Omega}^{\theta/(2r)}$ belongs to $L^{2r}_\omega$. 
   It is known that the $(\theta,r)$-R\'enyi divergences continue to satisfy the data processing inequality in the regime $r\in [1/2,1)$ and $r\geq \max(\theta,1-\theta)$ \cite{audenaert2015alpha}. In this range of parameters, the $(\theta,r)$-R\'enyi divergences are finite for arbitrary states of infinite systems. However, we will not attempt to prove the data processing inequality in this case. In matrix algebras, one can extend the range of the parameters to $\theta\in \mathbb{R} /\{1\}$ and $r>0$. The full range of parameters for which the $(\theta,r)$-R\'enyi divergence satisfies the data processing inequality was characterized in \cite{zhang2020wigner}.

\section{Multi-state measures}\label{sec:multistate}

We are now ready to generalize the construction of the two-state R\'enyi divergences to several states. For completeness, we have included a discussion of the H\"{o}lder inequality in the first subsection. The reader only interested in the multi-state R\'enyi divergences can skip this subsection.

\subsection{Generalized H\"{o}lder inequality}\label{sec:multiineq}

Consider the multi-state vector
\begin{eqnarray}\label{vecthetas}
   \ket{\Omega_{\vec{\psi}}(\vec{\theta},\vec{p} ) } =\Delta_{\Psi_1|\Omega}^{\theta_1/p_1}\cdots \Delta_{\Psi_n|\Omega}^{\theta_n/p_n} \ket{\Omega}
\end{eqnarray}
with $0\leq \theta_1+\cdots +\theta_n\leq 1$. We have introduced the compact notation $\vec{\theta}=(\theta_1,\cdots,\theta_n)$, $\vec{p}=(p_1,\cdots, p_n)$ and $\vec{\psi}=(\psi_1,\cdots, \psi_n)$. Note that by the relation (\ref{relmodeunitary}) the vector above only depends on the states $\omega_1$ to $\omega_n$ and not their purifications. We define the parameters $r_n$ and $p_{\vec{\theta}}$
\begin{eqnarray}
 &&\frac{1}{r_n}\equiv \frac{1}{p_1}+\cdots \frac{1}{p_n}\nn\\
&& \frac{1}{p_{\vec{\theta}}}\equiv \frac{\theta_1}{p_1}+\cdots +\frac{\theta_n}{p_n}\ .
\end{eqnarray}
We analytically continue the vector in (\ref{vecthetas}) to complex variables $z_i=\theta_i+it$. Since $p_{\vec{\theta}}\geq r_n$ the $r_n$-norm analytically continued to the complex strip is finite 
\begin{eqnarray}\label{frthetas}
 f_{\vec{\psi}|\omega}(\vec{z},\vec{p})=\|\ket{\Omega_{\vec{\psi}}(\vec{z},\vec{p})}\|_{r_n,\Omega}\ .
\end{eqnarray}
In matrix algebras, the function above is 
\begin{eqnarray}\label{frmatrix}
    &&f_{\vec{\psi}|\omega}(\vec{\theta},\vec{p})=\|\psi_1^{\theta_1/p_1}\cdots \psi_n^{\theta_n/p_n}\omega^{\frac{1}{r_n}-\frac{1}{p_{\vec{\theta}}}}\|_{r_n}\ .
    \end{eqnarray}
In what follows, we will use the fact that the function (\ref{frthetas}) is bounded and analytic on the complex domain of $\vec{z}$ with $0\leq \theta_1+\cdots+\theta_n\leq 1$ to prove the generalized H\"{o}lder inequality for the $(p,\omega)$-norms:\footnote{This was shown in theorem 5 of \cite{araki1982positive}.}
\begin{eqnarray}\label{thm5arakimasuda}
\|\Delta_{\Psi_1|\Omega}^{1/p_1}\cdots \Delta_{\Psi_n|\Omega}^{1/p_n}\ket{\Omega}\|_{r_n,\Omega}\leq \prod_{i=1}^n\|\Delta_{\Psi_i|\Omega}^{1/p_i}\ket{\Omega}\|_{p_i,\Omega}=\prod_{i=1}^n\|\psi_i\|^{1/p_i}_1\ .
\end{eqnarray}
Note that the measure above is independent of the state $\omega$.
If the states $\psi_i$ are all normalized the right-hand-side is equal to one. In matrix algebras, this is
\begin{eqnarray}
 \|\psi_1^{1/p_1}\cdots \psi_n^{1/p_n}\|_{r_n}\leq \prod_{i=1}^n\|\psi_i^{1/p_i}\|_{p_i}\ .
\end{eqnarray}
Defining the operators $a_i\equiv \psi_i^{1/p_i}$ gives the matrix form of the generalized H\"{o}lder inequality
\begin{eqnarray}
 \|a_1\cdots a_n\|_{r_n}\leq \|a_1\|_{p_1}\cdots \|a_n\|_{p_n}\ .
\end{eqnarray}
To prove (\ref{thm5arakimasuda}) we start by showing
\begin{eqnarray}
 \|\Delta_{\Psi_1|\Omega}^{1/p_1}\cdots \Delta_{\Psi_{n-1}|\Omega}^{1/p_{n-1}}\Delta_{\Psi_n|\Omega}^{1/p_n} &&\ket{\Omega}\|_{r_n,\Omega} \nn \\
 && \leq \|\Delta_{\Psi_1|\Omega}^{1/p_1}\cdots \Delta_{\Psi_{n-1}|\Omega}^{1/p_{n-1}}\ket{\Omega}\|_{r_{n-1},\Omega} \|\Delta_{\Psi_n|\Omega}^{1/p_n}\ket{\Omega}\|_{p_n,\Omega} \nn
\end{eqnarray}
for arbitrary $n$. Define 
\begin{eqnarray}
 &&\frac{1}{r_{n,\theta}}\equiv \frac{\theta}{r_{n-1}}+\frac{1-\theta}{p_n}
 %\nn\\
 %&&\frac{1}{p_{n,\theta}}=\frac{\theta}{r_{n-1}}+\frac{(1-\theta)}{p_n}\ .
\end{eqnarray}
and the function
\begin{eqnarray}\label{fvec}
 f_{\vec{\psi}|\omega}(\vec{\theta},\vec{p})\equiv\|\Delta_{\Psi_1|\Omega}^{\theta/p_1}\cdots \Delta_{\Psi_{n-1}|\Omega}^{\theta/p_{n-1}}\Delta_{\Psi_n|\Omega}^{(1-\theta)/p_n}\ket{\Omega}\|_{r_{n,\theta},\Omega}\ .
\end{eqnarray}
It can be analytically continued to complex $z=\theta+it$. 
%On the complex strip $\theta\in [0,1]$ it is bounded and holomorphic.
% , therefore by the Phragmen-Lindeloff principle (the maximum modulus principle applied to the strip) it takes it maximum value on the boundary the boundaries $\theta=0$ or $\theta=1$. 

Consider a general function $f(z)$ that is bounded and holomorphic in the complex strip $\theta\in [0,1]$ and continuous on its boundaries. Define the function $g(z)=f(z)f(0)^{z-1}f(1)^{-z}$ which is also holomorphic and bounded in the strip and continuous on the boundaries of the strip. The function $g(z)$ has value less than or equal to one on the boundaries, therefore by the Phragm\'{e}n-Lindel\"{o}f principle (the maximum modulus principle applied for the holomorphic functions bounded in the strip) it takes its maximum on the boundary. As a result, $|g(z)|\leq 1$ everywhere in the strip. On the real line $z=\theta$ we obtain the inequality
\begin{eqnarray}
&&|f(\theta)|\leq |f(0)|^{1-\theta} \:|f(1)|^{\theta}\ .
\end{eqnarray}
This result is sometimes called the Hadamard three-lines theorem.

Applying the argument above to our function in (\ref{fvec}) gives the inequality
\begin{eqnarray}
  f_{\vec{\psi}|\Omega}(\vec{\theta},\vec{p})\leq \|\Delta_{\Psi_1|\Omega}^{1/p_1}\cdots \Delta_{\Psi_{n-1}|\Omega}^{1/p_{n-1}}\ket{\Omega}\|^{\theta}_{r_{n-1},\Omega} \|\Delta_{\Psi_n|\Omega}^{1/p_n}\ket{\Omega}\|_{p_n,\Omega}^{1-\theta}\ .
\end{eqnarray}
Choosing $\theta=1/2$ and sending $p_i\to 2p_i$ gives
\begin{eqnarray}
 \|\Delta_{\Psi_1|\Omega}^{1/p_1}\cdots \Delta_{\Psi_n|\Omega}^{1/p_n} && \ket{\Omega}\|_{r_n,\Omega} \nn \\
 && \leq \|\Delta_{\Psi_1|\Omega}^{2/p_1}\cdots \Delta_{\Psi_{n-1}|\Omega}^{2/p_{n-1}}\ket{\Omega}\|^{1/2}_{r_{n-1}/2,\Omega} \|\Delta_{\Psi_n|\Omega}^{2/p_n}\ket{\Omega}\|^{1/2}_{p_n/2,\Omega}\nn\\
 &&=\|\Delta_{\Psi_1|\Omega}^{2/p_1}\cdots \Delta_{\Psi_{n-1}|\Omega}^{2/p_{n-1}}\ket{\Omega}\|^{1/2}_{r_{n-1}/2,\Omega}\|\psi_n\|^{1/p_n}_1\ .
\end{eqnarray}
Repeating this argument and using
\begin{eqnarray}
 \|\Delta_{\Psi|\Omega}^{1/p}\ket{\Omega}\|_{p,\Omega}=\|\psi\|^{1/p}_1
\end{eqnarray}
we obtain the generalized H\"{o}lder inequality in (\ref{thm5arakimasuda}).

\subsection{Three-state R\'enyi divergences}\label{sec:threestates}

In this subsection, we introduce the three-state R\'enyi divergences and use the monotonicity of the relative modular operator show that they satisfy the data processing inequality. 
For any operator monotone function $f$ with $f(1)=1$ and positive operators $X$ and $Y$ the Kubo-Ando mean $\sharp_f$ is defined to be \cite{kubo1980means,simon2019operator}
\begin{eqnarray}\label{kubdoando}
   X\sharp_f Y=X^{1/2}f(X^{-1/2}Y X^{-1/2})X^{1/2}
\end{eqnarray}
where we are assuming that $X$ is invertible. Note that $X\sharp_f X=X$. The most important properties of the Kubo-Ando mean for us are the monotonicity relation and the transformer inequality: 
\begin{enumerate}\label{Kubofprop}
    \item If $X_A\leq X_B$ and $Y_A\leq Y_B$ then $X_A\sharp_f Y_A\leq X_B\sharp_f Y_B$
\item For any $T$ we have 
\begin{eqnarray}\label{mapT}
T (X\sharp_f Y) T^\dagger\leq (T X T^\dagger)\sharp_f (T Y T^\dagger)
\end{eqnarray} with equality when $T$ is invertible.
\end{enumerate}
To simplify our equations we introduce the following notation:\footnote{
In what follows, we could have chosen a more general case 
\begin{eqnarray}
 \Delta_{\Psi_1,\Psi_2|\Omega}^f(g_1,g_2)\equiv g_1(\Delta_{\Psi_1|\Omega})\sharp_f g_2(\Delta_{\Psi_2|\Omega})
\end{eqnarray}
where $g_1$ and $g_2$ are arbitrary operator monotone functions such that such $g_i(x)\geq 0$ for $x\geq 0$, however, to keep the presentation clean we restrict to the operator monotone functions $g_1(x)=x^{\theta_1}$ and $g_2(x)=x^{\theta_2}$ as we did in (\ref{monoDeltaf}). The definition of the multi-state R\'enyi divergences generalizes in the straightforward way. Our proof of the data processing inequality will apply to this most general case.}
\begin{eqnarray}
 &&\Delta_{\Psi|\Omega;A}\equiv \Delta_{\Psi_A|\Omega_A}\nn\\
 &&\Delta^f_{\Psi_1,\Psi_2|\Omega}(\theta_1,\theta_2)\equiv   \Delta_{\Psi_1|\Omega}^{\theta_1}\sharp_f \Delta^{\theta_2}_{\Psi_2|\Omega}\ .
\end{eqnarray}
Choosing $\ket{\Omega}$ as the reference vector, $\ket{\Psi_1}$ and $\ket{\Psi_2}$ and $\theta\in (0,1)$ we have two monotonicity equations for the relative modular operators.
Combining these two inequalities using the Kubo-Ando mean and applying its property in (\ref{mapT}) we obtain
\begin{eqnarray}\label{monoDeltaf}
 F^\dagger \Delta^f_{\Psi_1,\Psi_2|\Omega;A}(\theta_1,\theta_2)F && \leq (F^\dagger \Delta^{\theta_1}_{\Psi_1|\Omega;A}F) \sharp_f (F^\dagger\Delta^{\theta_2}_{\Psi_2|\Omega;A}F)
  \nn \\ && \leq \Delta^f_{\Psi_1,\Psi_2|\Omega;B}(\theta_1,\theta_2)
 \ .
\end{eqnarray}
% In other words, 
% \begin{eqnarray}\label{contract3states}
%  F^f_{\theta_1,\theta_2}\equiv \lb \Delta^f_{\Psi_1,\Psi_2|\Omega;A}\rb^{1/2} F \lb\Delta^f_{\Psi_1,\Psi_2|\Omega;B} \rb^{-1/2}
% \end{eqnarray}
% is a contraction.
% We remind the reader that the $f$ in $\Delta^f_{\Psi_1,\Psi_2|\Omega}$ is convenient notation and has nothing to do with taking powers of the operator.
The first inequality becomes an equality when $F$ is invertible.
%  and
% \begin{eqnarray}
% %  &&\Delta^f_{\Psi,\Psi|\Omega}(\theta_1,\theta_2)=f(\Delta_{\Psi|\Omega}^{\theta_2-\theta_1})\Delta_{\Psi|\Omega}^{\theta_1}\nn\\
%  &&\Delta^f_{\Psi,\Psi|\Omega}(\theta,\theta)=\Delta_{\Psi|\Omega}^\theta\ .
%  \end{eqnarray}
 As before, we define the contraction
\begin{eqnarray}\label{Ff}
   F^f_{\theta_1,\theta_2}\equiv  \lb  \Delta^f_{\Psi_1,\Psi_2|\Omega;A}(\theta_1,\theta_2)\rb^{1/2}F\lb\Delta^f_{\Psi_1,\Psi_2|\Omega;B}(\theta_1,\theta_2)\rb^{-1/2}
\end{eqnarray}
and the three-state $f$-divergence
\begin{eqnarray}\label{threestateRenyi}
 S^f_{\theta_1,\theta_2}(\psi_1,\psi_2\|\omega)\equiv &&-2r\log\left\|\lb\Delta_{\Psi_1,\Psi_2|\Omega}^f(\theta_1/r,\theta_2/r)\rb^{1/2}\ket{\Omega}\right\|_{2r,\Omega}\nn\\
&&=-2r\log\left\|\lb \Delta_{\Psi_1|\Omega}^{\theta_1/r}\sharp_f\Delta_{\Psi_2|\Omega}^{\theta_2/r}\rb^{1/2}\ket{\Omega}\right\|_{2r,\Omega}
\end{eqnarray}
for $0\leq \theta_1,\theta_2\leq 1$, $r\in [1/2,\infty]$ and $f$ any operator monotone function with $f(1)=1$. It is clear from (\ref{relmodeunitary}) that the measure is independent of the purifications of $\psi_1$ and $\psi_2$. 
To prove the data processing inequality for this three-state measure,
we use the contraction in (\ref{Ff}) to write
\begin{eqnarray}
  \Big\| \Big( \Delta^f_{\Psi_1,\Psi_2|\Omega;A} &&(\theta_1,\theta_2) \Big)^{1/2} \ket{\Omega_A} \Big\|_{2r,\Omega_A} \nn \\
 &&=\Big\|F^f_{\theta_1,\theta_2} \lb\Delta^f_{\Psi_1,\Psi_2|\Omega;B}(\theta_1,\theta_2)\rb^{1/2}\ket{\Omega_B}\Big\|_{2r,\Omega_A}\nn\\
%  &&=\sup_{\|\ket{\chi}\|_{(2s,B)}=1}|\braket{W_\theta^\dagger\chi|\Delta^{\theta_1}_{\Psi_1|\Omega;B}\sharp_f \Delta^{\theta_2}_{\Psi_2|\Omega;B} |\Omega_B}|\nn\\
 &&\leq \Big\|F^f_{\theta_1,\theta_2} \Big\|_{(2r,\Omega_B)\to (2r,\Omega_A)}\Big\|\lb\Delta^f_{\Psi_1,\Psi_2|\Omega;B}(\theta_1,\theta_2)\rb^{1/2}\ket{\Omega_B}\Big\|_{2r,\Omega_B}\nn\\
 &&\leq \Big\|\lb \Delta^f_{\Psi_1,\Psi_2|\Omega;B}(\theta_1,\theta_2)\rb^{1/2}\ket{\Omega_B}\Big\|_{2r,\Omega_B}\ .
\end{eqnarray}
%where we have used  (\ref{invarfirst}) for the first equality.
This proves the data processing inequality for $r\geq 1$:
\begin{eqnarray}
 S^f_{\theta_1,\theta_2;r}(\psi_1,\psi_2\|\omega;B)\leq  S^f_{\theta_1,\theta_2;r}(\psi_1,\psi_2\|\omega;A)\ .
\end{eqnarray}
% If both $\theta_1$ and $\theta_2$ are negative from the integral representation for $\alpha\in (0,1)$
% \begin{eqnarray}
% X^{-\alpha}=\frac{\sin(\pi \alpha)}{\pi}\int_0^\infty dt\:\lb\frac{t^{-\alpha}}{t+X}\rb\ .
% \end{eqnarray}
% we have the two inequalities for $i=1,2$
% \begin{eqnarray}
%  F^\dagger \Delta_{\Psi_i|\Omega;A}^{-\theta_i}F\geq \Delta_{\Psi|\Omega;B}^{-\theta_i}
% \end{eqnarray}
% Then, we 

As a particular example, we choose $f(x)=x^\alpha$ with $\alpha=(0,1)$ as the operator monotone function. The Kubo-Ando geometric mean is 
\begin{eqnarray}\label{sharpalpha}
&&X\sharp_\alpha Y\equiv X^{1/2}\lb X^{-1/2}Y X^{-1/2}\rb^\alpha X^{1/2}
\end{eqnarray}
which satisfies the properties
\begin{enumerate}
    \item $(X_1\otimes X_2)\sharp_\alpha(Y_1\otimes Y_2)=(X_1\sharp_\alpha Y_1)\otimes (X_2 \sharp_\alpha Y_2)$
    \item If $[X,Y]=0$ then $X^{\theta_1}\sharp_\alpha Y^{\theta_2}=X^{(1-\alpha)\theta_1}Y^{\alpha\theta_2}$\ .
    % \item $X\sharp_\alpha 1 = X^{1-\alpha}$ and $1\sharp_\alpha X = X^{-\alpha}$.
\end{enumerate}
We define the three-state R\'enyi divergences
\begin{eqnarray}\label{threestateAlpha}
   &&S^{\alpha}_{\theta_1,\theta_2}(\psi_1,\psi_2\|\omega)  \equiv \frac{-2r}{(1-\theta_1)(1-\theta_2)}\log\left\|\lb\Delta^\alpha_{\Psi_1,\Psi_2|\Omega}(\theta_1,\theta_2;r)\rb^{1/2}\ket{\Omega}\right\|_{2r,\Omega}\nn\\
   &&\Delta^\alpha_{\Psi_1,\Psi_2|\Omega}(\theta_1,\theta_2;r)\equiv \Delta_{\Psi_1|\Omega}^{\frac{\theta_1}{(1-\alpha)r}}\sharp_\alpha \Delta_{\Psi_2|\Omega}^{\frac{\theta_2}{\alpha r}}\ .
\end{eqnarray}
Note that $\alpha$ in $\Delta^\alpha_{\Psi_1,\Psi_2|\Omega}$ is simply an index and not a power. The powers of the relative modular operator are chosen such that when the relative modular operators commute the measure is independent of $\alpha$. In matrix algebras, this measure is
\begin{eqnarray}\label{threestatealphamatrix}
    S^{\alpha}_{\theta_1,\theta_2}(\psi_1,\psi_2\|\omega)  &&\equiv \frac{-2r}{(1-\theta_1)(1-\theta_2)}\log\left\|\lb\psi_1^{\frac{\theta_1}{(1-\alpha)r}}\sharp_\alpha \psi_2^{\frac{\theta_2}{\alpha r}}\rb^{1/2} \omega^{\frac{\theta_0}{2r}}\right\|_{2r}
\end{eqnarray}
where $\theta_0+\theta_1+\theta_2=1$. 

\paragraph{Special cases:} In the $\theta_0\to 0$, the expression above is independent of $\omega$ and we obtain
 \begin{eqnarray}
  S^\alpha_{1-\theta,\theta;r}(\psi_1,\psi_2\|\omega)=\frac{r}{\theta(\theta-1)}\log\left\|\psi_1^{\frac{(1-\theta)}{(1-\alpha)r}}\sharp_\alpha \psi_2^{\frac{\theta}{r\alpha}}\right\|_{r}\ .
\end{eqnarray}
If we further set $\alpha=\theta$, up to an overall coefficient, it reduces to a generalization of the geometric divergence defined in \cite{matsumoto2015new}:
\begin{eqnarray}
 S^\theta_{1-\theta,\theta;r}(\psi_1,\psi_2\|\omega)=\frac{r}{\theta(\theta-1)}\log \left\|\psi_1^{\frac{1}{r}}\sharp_\theta \psi_2^{\frac{1}{r}}\right\|_{r}\ .
\end{eqnarray}

In the special cases 
 $\theta_1\to 0$ (or $\theta_2\to 0$), the three-state measure in (\ref{threestatealphamatrix}) reduces to the $(\theta,r)$-R\'enyi divergence
\begin{eqnarray}
&&S^\alpha_{0,\theta;r}(\psi_1,\psi_2\|\omega)=S_{\theta,r}(\psi_2\|\omega)\nn\\
&&S^\alpha_{\theta,0;r}(\psi_1,\psi_2\|\omega)=S_{\theta,r}(\psi_1\|\omega)\ .
\end{eqnarray}
Another special case where we recover the $(\theta,r)$-R\'enyi divergence is $\psi_1=\psi_2$:
\begin{eqnarray}
   S_{\theta_1,\theta_2;r}(\psi,\psi\|\omega)&&=\frac{-2r}{(\theta_1-1)(\theta_2-1)}\log\|\psi^{(\theta_1+\theta_2)/(2r)}\omega^{\theta_0/(2r)}\|_{2r}\nn\\
   &&=\frac{\theta_0}{(\theta_1-1)(\theta_2-1)}S_{\theta_1+\theta_2,r}(\psi\|\omega)\ .
\end{eqnarray}
When $\alpha=1/2$ it is convenient to introduce the notation
 \begin{eqnarray}\label{sharp}
  X\sharp Y=X^{1/2}\lb X^{-1/2}Y X^{-1/2}\rb^{1/2}X^{1/2}\ .
 \end{eqnarray}
to write
 \begin{eqnarray}
  S^{1/2}_{\theta_1,\theta_2;r}(\psi_1,\psi_2\|\omega)=\frac{-2r}{(1-\theta_1)(1-\theta_2)}\log\left\|\lb \psi_1^{\theta_1/r}\sharp \psi_2^{\theta_2/r}\rb^{1/2} \omega^{\theta_0/(2r)}\right\|_{2r}\ .
 \end{eqnarray}

\subsection{Multi-state R\'enyi divergences}

The generalization to arbitrary number of states is straightforward.  %but the following arguments work for arbitrary $\sharp_f$. 
We use the vector notation $\vec{\Psi}=(\Psi_1,\cdots, \Psi_n)$, $\vec{\theta}=(\theta_1,\cdots ,\theta_n)$ and $\vec{f}=(f_1,\cdots, f_{n-1})$ to define the operator
\begin{eqnarray}\label{DeltaThetas}
   \Delta^{\vec{f}}_{\vec{\Psi}|\Omega}(\vec{\theta})\equiv \Delta_{\Psi_1|\Omega}^{\theta_1}\sharp_{f_1}\cdots \sharp_{f_{n-1}}\Delta_{\Psi_n|\Omega}^{\theta_n}\ .
 \end{eqnarray}
 We are using the simplified notation\footnote{Multi-variate operator geometric means were discussed in \cite{sagae1994upper}.}
 \begin{eqnarray}
  X_1\sharp_{f_1} X_2\sharp_{f_2} X_3\equiv X_1\sharp_{f_1} (X_2\sharp_{f_2} X_3)\ .
 \end{eqnarray}
 We define the multi-state $f$-divergence to be
 \begin{eqnarray}\label{multistateRenyif}
  S^{\vec{f}}_{\vec{\theta};r}(\vec{\psi}\|\omega)=\frac{ -2r}{\prod_{i=1}^n(1-\theta_i)}\log\left\|\lb\Delta^{\vec{f}}_{\vec{\Psi}|\Omega}(\vec{\theta})\rb^{1/2}\ket{\Omega}\right\|_{2r,\Omega}\ .
 \end{eqnarray}
 This is a special case of the more general measure
 \begin{eqnarray}
  &&S^{\vec{f},\vec{g}}_r(\vec{\psi}\|\omega)=\frac{-1}{N(\vec{g})} \log \left\|\lb\Delta^{\vec{f}}_{\vec{\Psi}|\Omega}(\vec{g})\rb^{1/2}\ket{\Omega}\right\|_{2r,\Omega}\nn\\
  &&\Delta^{\vec{f}}_{\vec{\Psi}|\Omega}(\vec{g})\equiv g_1(\Delta_{\Psi_1|\Omega})\sharp_{f_1}\cdots \sharp_{f_{n-1}} g_n(\Delta_{\Psi_n|\Omega})
 \end{eqnarray}
 for operator monotone functions $f_1,\cdots, f_{n-1}$ with $f_i(1)=1$ and $g_1, \cdots,g_n$ with $g_i$ satisfying $g_i(x)\geq 0$ for all $x\geq 0$. Moreover, $\frac{-1}{N(\vec{g})}$ is a normalization. In the remainder of this work, we focus on the measure in (\ref{multistateRenyif}). We will see that when $\theta_1+\cdots \theta_n=1$ this measure is independent of $\ket{\Omega}$.
 
 To prove the data processing inequality, as before, we first construct the inequality
 \begin{eqnarray}
   F^\dagger \Delta^{\vec{f}}_{\vec{\Psi}|\Omega;A}(\vec{\theta})F\leq \Delta^{\vec{f}}_{\vec{\Psi}|\Omega;B}(\vec{\theta})
 \end{eqnarray}
 by repeatedly using (\ref{monoDeltaf}), from which we get the contraction
 \begin{eqnarray}
     F^{\vec{f}}_{\vec{\theta}}\equiv  \lb \Delta^{\vec{f}}_{\vec{\Psi}|\Omega;A}(\vec{\theta})\rb^{1/2}F\lb \Delta^{\vec{f}}_{\vec{\Psi}|\Omega;B}(\vec{\theta})\rb^{-1/2}\ .
 \end{eqnarray}
%  \begin{eqnarray}
%   \Delta_{\vec{\psi}|\Omega;B}(\vec{\theta})\geq W \Delta_{\vec{\psi}|\Omega;A}(\vec{\theta}) W^\dagger\ . 
%  \end{eqnarray}
We have
\begin{eqnarray}
 \left\|\lb\Delta^{\vec{f}}_{\vec{\Psi}|\Omega;A}(\vec{\theta})\rb^{1/2}\ket{\Omega_A}\right\|_{2r,\Omega_A}&&=\left\|F^{\vec{f}}_{\vec{\theta}}\lb\Delta^{\vec{f}}_{\vec{\Psi}|\Omega;B}(\vec{\theta})\rb^{1/2}\ket{\Omega_B}\right\|_{2r,\Omega_A}\nn\\
 &&\leq \|F^{\vec{f}}_{\vec{\theta}}\|_{(2r,\Omega_B)\to (2r,\Omega_A)}\left\|\lb\Delta^{\vec{f}}_{\vec{\Psi}|\Omega;B}(\vec{\theta})\rb^{1/2}\ket{\Omega_B}\right\|_{2r,\Omega_B}\nn\\
 && \leq \left\|\lb\Delta^{\vec{f}}_{\vec{\Psi}|\Omega;B}(\vec{\theta})\rb^{1/2}\ket{\Omega_B}\right\|_{2r,\Omega_B} .
\end{eqnarray}
This implies that the multi-state $f$-divergences satisfy the data processing inequality for $r\geq 1$
\begin{eqnarray}
 S^{\vec{f}}_{\vec{\theta},r}(\vec{\psi_B}\|\omega_B)\leq S^{\vec{f}}_{\vec{\theta},r}(\vec{\psi_A}\|\omega_A)
\end{eqnarray}
for any quantum channel $\Phi^*$.

To be more concrete, we restrict to the geometric mean $\sharp_\alpha$ in (\ref{sharpalpha}). Consider $n$ operators $X_1$ to $X_n$ that pairwise commute. Define $\alpha_n=\alpha_0=0$ so that 
\begin{eqnarray}\label{gammai}
 &&X_1^{\theta_1}\sharp_{\alpha_1}\cdots \sharp_{\alpha_{n-1}}X_n^{\theta_n}=X_1^{\gamma_1\theta_1}\cdots X_n^{\gamma_n\theta_n}\nn\\
 &&\gamma_i=(1-\alpha_i)(\alpha_1\cdots \alpha_{i-1})\ .
\end{eqnarray}
Note that $\gamma_i$ are all positive and add up to one, hence, they are a probability distribution.
We define the operator
\begin{eqnarray}
 \Delta_{\vec{\Psi}|\Omega}^{\vec{\alpha}}(\vec{\theta};r)\equiv \Delta_{\Psi_1|\Omega}^{\frac{\theta_1}{r\gamma_1}}\sharp_{\alpha_1}\cdots \sharp_{\alpha_{n-1}}\Delta_{\Psi_n|\Omega}^{\frac{\theta_n}{r\gamma_n}}\ .
\end{eqnarray}
The advantage of this definition is that it is independent of $\vec{\alpha}$ when the relative modular operators commute.
Then, the multi-state R\'enyi divergence is
 \begin{eqnarray}\label{multistateRenyialpha}
  S^{\vec{\alpha}}_{\vec{\theta};r}(\vec{\psi}\|\omega)=\frac{ -2r}{\prod_{i=1}^n(1-\theta_i)}\log\left\|\lb\Delta^{\vec{\alpha}}_{\vec{\Psi}|\Omega}(\vec{\theta};r)\rb^{1/2}\ket{\Omega}\right\|_{2r,\Omega}\ .
 \end{eqnarray}
 %where $\vec{\theta}/r\equiv (\theta_1/r,\cdots, \theta_n/r)$.
% We define $\alpha_n=\alpha_0=0$ and write the identity
% \begin{eqnarray}
%  &&X^{\theta_1}\sharp_{\alpha_1}\cdots \sharp_{\alpha_{n-1}}X^{\theta_n}=X^{\gamma_0}\nn\\
%  &&\gamma_0=\sum_{i=1}^n \theta_i(1-\alpha_i)(\alpha_1\cdots \alpha_{i-1})
% \end{eqnarray}
In matrix algebras, this measure becomes
%we have the multi-state R\'enyi divergences 
\begin{eqnarray}\label{multiRenyi}
 S^{\vec{\alpha}}_{\vec{\theta},r}(\vec{\psi}\|\omega)=\frac{-2r}{\prod_{i=1}^n(1-\theta_i)}\log \left\|\lb\psi_1^{\frac{\theta_1}{r\gamma_1}}\sharp_{\alpha_1}\cdots \sharp_{\alpha_{n-1}}\psi_n^{\frac{\theta_n}{r\gamma_n}}\rb^{1/2}\omega^{\frac{\theta_0}{2r}}\right\|_{2r}
\end{eqnarray}
where $\theta_0+\theta_1+\cdots +\theta_n=1$. We can think of $\theta_i$ as a probability distribution associated with states $\psi_i$. As before, when $\theta_0=0$ the measure above is independent of $\omega$. 

Similar to (\ref{negativetheta}) we can divide our multi-state R\'enyi measure by $(1-\theta_0)$ to make it more symmetric among $\theta_0$ and the rest of $\theta_i$:
\begin{eqnarray}\label{moresymm}
\hat{S}^{\vec{\alpha}}_{\vec{\theta},r}(\vec{\psi}\|\omega)\equiv \frac{1}{1-\theta_0}S^{\vec{\alpha}}_{\vec{\theta},r}(\vec{\psi}\|\omega)\ .
\end{eqnarray}

\paragraph{Special cases:}
In the limit $r\to \infty$, we have the multi-variate Lie-Trotter formula for self-adjoint operators $a_1,\cdots, a_n$ \cite{bhatia2013matrix,sutter2017multivariate}
\begin{eqnarray}
 \lim_{r\to \infty}\lb e^{a_1/r}\cdots e^{a_n/r}\rb^r=e^{a_1+\cdots +a_n}\ .
\end{eqnarray}
In lemma 3.3 of \cite{hiai1993golden} it was shown that for $\alpha\in [0,1]$ and $a_1$ and $a_2$ self-adjoint
\begin{eqnarray}
 \lim_{r\to \infty}\lb e^{a_1/r}\sharp_\alpha e^{a_2/r}\rb^r=e^{(1-\alpha)a_1+\alpha a_2}\ .
\end{eqnarray}
This was further generalized by \cite{ahn2007extended} to multi-variate geometric means
\begin{eqnarray}
 &&\lim_{r\to \infty}\lb e^{a_1/r}\sharp_{\alpha_1}\cdots \sharp_{\alpha_{n-1}}e^{a_{n-1}/r}\rb^r=e^{\sum_i  \gamma_i a_i}
%  \nn\\
%  &&\beta_i=(1-\alpha_i)(\alpha_1\cdots \alpha_{i-1})
\end{eqnarray}
with $\gamma_i$ given in (\ref{gammai}). Notice that the right-hand-side of the equation above is invariant under the permutations of $a_i$. Applied to our measure, we find
\begin{eqnarray}
 \lim_{r\to \infty}S^{\vec{\alpha}}_{\vec{\theta},r}(\vec{\psi}\|\omega)&&=\frac{-1}{(1-\theta_1)\cdots (1-\theta_n)}\log\tr\lb e^{\sum_i  \theta_i\log \psi_i+\theta_0\log \omega}\rb
\end{eqnarray}
which is independent of $\alpha_i$. Now, except for an overall $1/(1-\theta_0)$ factor, the reference state $\omega$
is no longer distinguished from the rest. 
We include $\omega$ inside $\vec{\psi}$ as $\psi_0$. We define the vector $\vec{\theta_\epsilon}$ that is $\theta_j=1-\epsilon$ for a particular $j$, and $\theta_i=\epsilon \beta_i$ for $i\neq j$ including $\theta_0=\epsilon \beta_0$. Since $\vec{\theta}_\epsilon$ is a probability distribution the weights $\beta_i$ sum up to one; hence $\beta_i$ is also a probability distribution. In the limit $\epsilon\to 0$, all $\theta_i\to 0$ except for $\theta_j$ that goes to one and we find\footnote{Since the measure does not depend on $\vec{\alpha}$ we suppress it in the notation.}
\begin{eqnarray}\label{epsilonzero}
 \lim_{\epsilon \to 0}S_{\vec{\theta}_\epsilon,\infty}(\vec{\psi})=\sum_{i=0}^n\beta_i\tr\lb \psi_j(\log \psi_j-\log \psi_i)\rb=\sum_{i=0}^{n}\beta_i S(\psi_i\|\psi_j)
\end{eqnarray}
which is the weighted average of the relative entropies of $\psi_i$ with respect to $\psi_j$.

The same analysis can be repeated at finite $r$ if all the states commute. In this case, we have $n$ probability distributions and our multi-state measure is independent of both $r$ and the vector $\vec{\alpha}$:
\begin{eqnarray}
 D_{\vec{\theta}}(\{p_1\},\cdots,\{p_n\})=\frac{-1}{(1-\theta_1)\cdots (1-\theta_n)}\log\lb\sum_{x\in X}p_1(x)^{\theta_1}\cdots p_n(x)^{\theta_n}\rb\ .
\end{eqnarray}
This is the generating functional in (\ref{generating}). Taking the same $\epsilon\to 0$ limit of $\vec{\theta_\epsilon}$ gives a weighted average of the relative entropies:
\begin{eqnarray}
 \lim_{\epsilon\to 0}D_{\vec{\theta_\epsilon}}(\vec{p})=\sum_{i=1}^n \beta_i D_{KL}(p_i\|p_j)\ .
\end{eqnarray}

Consider the the multi-state measure in \ref{moresymm}.
In appendix \ref{app:theta}, we show that in case where we set  $\theta_i=\epsilon \beta_i$ and $\theta_0=1-\epsilon$, at finite $r$, we obtain the same weighted average of relative entropies:
\begin{eqnarray}
 \lim_{\epsilon\to 0}\hat{S}^{\vec{\alpha}}_{\vec{\theta_\epsilon};r}(\vec{\psi}\|\omega)=\sum_{i=1}^n \beta_i S(\psi_i\|\omega)\ .
\end{eqnarray}

\section{Infinite dimensions}\label{sec:infinite}

In this section, we generalize our discussion of $L^p_\omega$ spaces and the multi-state R\'enyi divergences to an arbitrary von Neumann algebra. . This includes the local algebra of quantum field theory (QFT) that is a type III algebra, meaning that it does not admit a trace.\footnote{Formally, a trace is a normal completely positive (CP) map from the algebra to the complex numbers $\tr:\mA\to \mathbb{C}$ that satisfies 
$\forall a_1,a_2\in \mA:\quad \tr(a_1a_2)=\tr(a_2a_1)$.} We closely follow the reference \cite{araki1982positive}.

Any normal CP map $\omega:\mA\to \mathbb{C}$ that satisfies $\omega(1)=1$ is called a state. In infinite dimensions, the vector $\ket{e}$ or a trace might not exist. However, we can use any normal state $\omega$ to define an inner product for the map $a\to a\ket{\Omega}$: 
\begin{eqnarray}
\braket{a_1\Omega|a_2\Omega}=\omega(a_1^\dagger a_2)\ . 
\end{eqnarray} 
The closure of the set  $a\ket{\Omega}$ is the GNS Hilbert space $\mH_\omega$. For simplicity, we have restrict to the case of faithful normal states.

The Tomita operator $S_\Omega:\mH_\omega\to \mH_\omega$ is the anti-linear operator defined by
\begin{eqnarray}
S_\Omega a\ket{\Omega} =a^\dagger \ket{\Omega}\ .
\end{eqnarray}
The closure of $S_\Omega$ has a polar decomposition
\begin{eqnarray}
 S_\Omega=J_\Omega \Delta_\Omega^{1/2}
\end{eqnarray}
where $J_\Omega$ and $\Delta_{\Omega}=\Delta_{\Omega|\Omega}$ are the generalizations of the modular conjugation and the modular operator to arbitrary von Neumann algebras.
The natural cone is the set of vectors that are invariant under $J_\Omega$. The vectors in the natural cone are in one-to-one correspondence with the normal states on $\mA$. The relative Tomita operator is defined by the equation
\begin{eqnarray}
 S_{\Psi|\Omega}a\ket{\Omega}=a^\dagger\ket{\Psi}
\end{eqnarray}
with polar decomposition (after closure) 
\begin{eqnarray}
 S_{\Psi|\Omega}=J_{\Psi|\Omega}\Delta_{\Psi|\Omega}^{1/2},
\end{eqnarray}
where $\Delta_{\Psi|\Omega}$ is the generalization of the relative modular operator, and $J_{\Psi|\Omega}$ is an anti-unitary operator if both $\omega$ and $\psi$ are faithful. When $\ket{\Psi}$ belongs to the natural cone we have $J_{\Omega|\Psi}=J_\Omega$, otherwise $J_{\Omega|\Psi}J_\Omega$ is a partial isometry in $\mA'$; see \cite{haag2012local}.

Motivated by the expression (\ref{arakimasuda}) we define the $(p,\Omega)$-norm of a vector $\ket{\Psi}\in \mH_\omega$ as
\begin{eqnarray}
 &&\|\ket{\Psi}\|_{p,\Omega}=\sup_{\ket{\chi}\in \mH_\omega}\|\Delta_{\chi|\Omega}^{1/2-1/p}\ket{\Psi}\|,\qquad\forall p\in [2,\infty]\nn\\
 &&\|\ket{\Psi}\|_{p,\Omega}=\inf_{\ket{\chi}\in \mH_\omega}\|\Delta_{\chi|\Omega}^{1/2-1/p}\ket{\Psi}\|,\qquad\forall p\in [1,2)\ .
\end{eqnarray}
For $p\geq 2$ the $(p,\Omega)$-norm is finite if $\ket{\Psi}$ is in the intersection of the domains of $\Delta_{\chi|\Psi}^{1/2-1/p}$ for all $\ket{\chi}\in \mH_\omega$. When $\ket{\Psi}$ is outside of this intersection set we say $\|\ket{\Psi}\|_{p,\Omega}=\infty$. The closure of the set of all $\ket{\Psi}$ with finite $(p,\Omega)$-norm is called the $L^p_\omega$ space \cite{araki1982positive}.
For $p\in [1,2)$ the $L^p_\omega$ space is defined to be the completion of the Hilbert space $\mH_\omega$ with the $(p,\Omega)$-norm. In general, we have $L^p_\omega\subseteq L^r_\omega$ for $r\leq p$ and $L^\infty_\omega$ is the algebra itself with its operator norm $\|a\|_\infty$. The $L^2_\omega$ is the GNS Hilbert space $\mH_\omega$ and the $L^1_\omega$ is the space of normal linear functionals of $\mA$. 
We can embed the vectors $\ket{\Psi}\in\mH_\omega$ in $L^1_\omega$ using the map
\begin{eqnarray}
 \psi(\cdot)=\braket{\Psi|\cdot\Omega}\ .
\end{eqnarray}
However, since $L^1_\omega$ is larger than $\mH_\omega$ not all states $\psi$ can be expressed this way.

The $L^p_\omega$ space is dual to the $L^q_\omega$ space when $q$ is the H\"{o}lder dual of $p$:
\begin{eqnarray}
 \|\ket{\Psi}\|_{p,\omega}=\sup_{\|\ket{\chi}\|_{q,\omega}=1}|\braket{\chi|\Psi}|\ .
\end{eqnarray}
Given a normal state $\psi\in L^1_\omega$ the vector
\begin{eqnarray}
 \Delta_{\Psi|\Omega}^{1/p}\ket{\Omega}\in L^p_\omega
\end{eqnarray}
for $p\in [2,\infty)$.
% If $\psi$ does not belong to $L^2_\omega$ in the expression above the vector $\ket{\Psi}\notin \mH_\omega$. Instead, it is the vector in the GNS Hilbert space of $\psi$. 
For every vector $\ket{\chi}\in L^p_\omega$ there exists a unique $\psi\in L^1_\omega$ such that
\begin{eqnarray}
 \ket{\chi}=u \Delta_{\Psi|\Omega}^{1/p}\ket{\Omega}
\end{eqnarray}
with some partial isometry $u\in \mA$. The vector 
\begin{eqnarray}
 \ket{\Omega(\theta)}=\Delta_{\Psi|\Omega}^{\theta/2}\ket{\Omega}
\end{eqnarray}
is analytic in the complex strip $z=\theta+it$ with $\theta\in [0,1]$. The reason is that we can write
\begin{eqnarray}
&&\Delta_{\Psi|\Omega}^{\theta+it}\ket{\Omega}=\Delta_{\Psi|\Omega}^\theta (D\Psi:D\Omega)_t\ket{\Omega}
\end{eqnarray}
where 
\begin{eqnarray}
&& (D\Psi:D\Omega)_t\equiv\Delta_{\Psi|\Omega}^{it}\Delta_\Omega^{-it}\in \mA
\end{eqnarray}
is the Connes cocycle which is a partial isometry in the algebra for all real values of $t$ \cite{connes1973classification}.

All the multi-state measures discussed in the previous section and the inequalities they satisfy generalize to arbitrary von Neumann algebras except for (\ref{rinf}).\footnote{We do not know how to prove a generalization of (\ref{Srsym}) to arbitrary von Neumann algebras.}

\section{Quantum state discrimination}\label{sec:discrimination}

In asymmetric quantum state discrimination, we are given a state $\omega$ that we do not know a priori. The task is to perform measurements on this state to decide whether it is $\omega$ or any of the alternate hypotheses $K=\{\psi_0,\cdots \psi_k\}$. We would like to know what is the optimal measurement to perform on the state to make the decision and what is the minimum probability of misidentifying the state.
%This is sometimes called the asymmetric hypothesis testing.

First, consider the case with only one alternate hypothesis $\psi$.
Assume we are given $n$ identical copies of the state prepared in the form $\omega^{\otimes n}$ and we are allowed to use any measurement in the $n$-copy Hilbert space to identify the state. Denote by $\beta_n$ the probability that we misidentify the state as $\psi$ with the optimal measurement. Any other measurement strategy to distinguish the two states fails with probability larger than $\beta_n$. According to quantum Stein's lemma $\beta_n$ behaves asymptotically as \cite{hiai1991proper}
\begin{eqnarray}\label{stein}
   \lim_{n\to \infty} -\frac{1}{n}\log\beta_n=S(\psi\|\omega)\ .
\end{eqnarray}
This provides an operational interpretation for relative entropy. The asymmetry of the relative entropy is related to the fact that we assumed that in reality the state was $\omega$. Of course, if we were given the state $\psi$ instead the asymptotic error rates are controlled by $S(\omega\|\psi)$. In general, in hypothesis testing we have two types of errors and their corresponding optimal probabilities
\begin{enumerate}
    \item $\alpha_n$: the state was $\psi$ and we misidentified it as $\omega$.
    \item $\beta_n$: the state was $\omega$ and we misidentified it as $\psi$.
\end{enumerate}
There is a trade-off between these two types of errors.
Since we do not know whether the state is $\omega$ or $\psi$ we should try to adopt a strategy that minimizes a combination of both errors. One might expect that these strategies would fail with minimal probabilities that interpolate between $S(\psi\|\omega)$ and $S(\omega\|\psi)$ as we go from minimizing the type 2 to type 1 errors. This intuition is confirmed in symmetric hypothesis testing when we choose to minimize the average of the two error probability types. According to the quantum Chernoff bound, the optimal error probability in the symmetric case in the $n\to \infty$ limit is \cite{audenaert2007discriminating}
\begin{eqnarray}
   &&E_{e,n}\leq e^{-n C(\psi,\omega)}\nn\\
   &&C(\psi,\omega)=-\log \inf_{\theta\in (0,1)}\tr\lb \psi^{\theta}\omega^{1-\theta}\rb\ .
\end{eqnarray}
Note that the quantity $C(\psi,\omega)$ is related to a minimization over the Petz divergences in (\ref{Petzdiv}).
The in-between strategies succeed with probabilities that depend on the Petz divergences.
For instance, let us restrict to the measurements that leads to type 2 errors smaller than some constant $e^{-n r}$, i.e. $\beta_n\leq e^{-n r}$, and denote by $\alpha_{n,r}$ the optimal probability of the type 1 errors among these measurements. In the limit $n\to \infty$ we have \cite{mosonyi2015quantum}
\begin{eqnarray}\label{Hoeff}
   &&\alpha_{n,r}\leq e^{-n H_r(\psi\|\omega)}\nn\\
   &&H_r(\psi\|\omega)=\sup_{\theta\in (0,1)}\frac{\theta-1}{\theta}(r-D_\theta(\psi\|\omega))\ .
\end{eqnarray}
The quantity $H_r(\psi\|\omega)$ is called the Hoeffding divergence.
 The inequality above provides an operational interpretation for the Petz divergences $D_\theta(\psi\|\omega)$. 
It follows from (\ref{stein}) that if $r> S(\psi\|\omega)$ the error $\alpha_{n,r}$ tends to one exponentially fast for large $n$. It was shown in \cite{mosonyi2015quantum} that as $n \to \infty$
\begin{eqnarray}\label{converseHoeff}
   &&1-\alpha_{n,r}\leq e^{-n H^*_r(\psi\|\omega)}\nn\\
   &&H_r^*(\psi\|\omega)=\sup_{\theta>1} \frac{\theta-1}{\theta}(r-S_\theta(\psi\|\omega))\ .
\end{eqnarray}
The function $H^*_r(\psi\|\omega)$ is often called the converse Hoeffding divergence. It provides an operational interpretation for the sandwiched R\'enyi divergences.

Now, let us consider the completely asymmetric case where we are given $\omega$ but we have several alternate hypotheses $K=\{\psi_1,\cdots ,\psi_k\}$. The generalization of the quantum Stein's lemma in (\ref{stein}) to the multi-state setting is called the quantum Sanov's lemma \cite{bjelakovic2005quantum,hayashi2002optimal}. It says that given $\omega$ the optimal probability $\beta_n$ of mistaking it for other states at large $n$ is
\begin{eqnarray}
   &&\beta_n\leq e^{-n S(K\|\omega)}\nn\\
   &&S(K\|\omega)=\min_{\psi_i \in K}S(\psi_i\|\omega)\ .
\end{eqnarray}
In the symmetric case, given a set of hypothesis $K$, the multi-state Chernoff bound says that the 
minimal errors are controlled by the multi-state Chernoff distance \cite{li2016discriminating}
\begin{eqnarray}
 &&E_{e,n}\leq e^{-n \xi}\nn\\
 &&\xi=\min_{i\neq j}C(\psi_i,\psi_j)\ .
\end{eqnarray}
However, away from the asymmetric case when we have to minimize various types of errors that generalize the type 1 and type 2 errors to multi-state setting, one expects that the multi-state measures that control the optimal probabilities to interpolate between the relative entropies $S(\psi_i\|\omega)$ and $C(\psi_i,\psi_j)$. 
The optimal error probabilities satisfy a data processing inequality because all distinguishability measures are non-increasing as we restrict the set of allowed measurements. Our multi-state measures interpolate in between these measures as we vary the probability measure $(\theta_0,\theta_1,\cdots, \theta_m)$ and satisfy the data processing inequality.
We take this as an evidence to conjecture that the multi-state R\'enyi divergences in (\ref{multiRenyi}) have operational interpretations in asymmetric multi-state discrimination where we are given the state $\omega$ and the hypotheses are the states $\psi_1, \cdots, \psi_m$. 
One attempt to make this conjecture more precise is as follows:\footnote{We thank Milan Mosonyi for the suggestion.} In the multi-state setting with $m$ alternative hypotheses $\{\psi_1, \cdots, \psi_m\}$ there are $m$ probability errors $\beta_{i,n}$ associated with misidentifying $\omega$ with $\psi_i$. Choose a specific $j$ and restrict to the measurements with error probabilities $\beta_{i,n}\leq e^{-nr_i}$ for $i\neq j$ at large number of measurements $n$. One might expect that the optimal error probability for $j$ is given by an infimum over $\theta_i$ of some function of $r_i$ minus our multi-state measures. However, we do not know what function of $r_i$ is relevant or how to fix the value of the $\alpha_i$ parameters. In the classical limit, the $\alpha_i$ parameters go away making it easier to find the appropriate function of $r_i$, however we will not attempt that here.
For more recent developments in quantum state discrimination see \cite{mosonyi2020error,brandao2020adversarial}.

\section{Discussion}

In this work, we constructed multi-state R\'enyi divergences and proved that they satisfy the data processing inequality in the range $r\geq 1$ and $\theta_i\in [0,1]$. 
Both the Petz and the sandwiched R\'enyi divergences are monotonic in $p$; however, we did not explore potential monotonicity of our multi-state R\'enyi divergences in any of the parameters $r$ or $\theta$. 
We postpone this question to future work. 

Recently, Fawzi and Fawzi used the Kubo-Ando geometric to define new quantum R\'enyi divergences in terms of a convex optimization program and proved that they satisfy the data processing inequality \cite{fawzi2021defining}. It would be interesting to use the non-commutative $L^p_\omega$ spaces to rewrite their expressions as $(p,\omega)$-norms and explore their potential multi-state generalizations.
 
In section \ref{sec:multiineq} we analytically continued the vector (\ref{vecthetas}) to complex $\theta_i$. Consider the vectors $\ket{\Omega_i}=u_i\ket{\Omega}$ where $u_i\in \mA$ are unitary operators. In that case, the relative modular operator can be written in terms of the modular operator of $\omega$:
\begin{eqnarray}
 \Delta_{u\Omega|\Omega}=u \Delta_\Omega u^\dagger
\end{eqnarray}
where $\Delta_\Omega$ is the modular operator of $\Omega$.
Then, our analytically continued vector is
\begin{eqnarray}
 \ket{\Omega_{u_1,\cdots u_n}(\vec{z})}=u_1 \Delta_\Omega^{z_1}(u_1^\dagger u_2)\Delta_\Omega^{z_2}(u_2^\dagger u_3)\cdots \Delta_\Omega^{z_n}u_n^\dagger\ket{\Omega}\ .
\end{eqnarray}
If we take all $z_i$ to be imaginary we end up with modular evolved operators 
\begin{eqnarray}
 &&\|\ket{\Omega_{u_1,\cdots,u_n}(i\vec{t})}\|=\|(u_1^\dagger u_2)_{t_1}(u_2^\dagger u_3)_{t_1+t_2}\cdots u_n\ket{\Omega}\|\nn\\
 &&a_t\equiv \Delta_\Omega^{it}a\Delta_\Omega^{-it}\ .
\end{eqnarray}
For general values of $t_i$ we obtain a $2n$-point modular correlation function that is not modular time-ordered. 
In fact, since $a\in \mA$ belong to $L^\infty_\omega$ we can generalize our vector in (\ref{vecthetas}) by introducing operators $a_i\in \mA$ (not necessarily unitaries)
\begin{eqnarray}
 \Delta_{\Psi_1|\Omega}^{z_1}a_1\cdots \Delta_{\Psi_n|\Omega}^{z_n}a_n\ket{\Omega}\ .
\end{eqnarray}
setting $\ket{\Omega_i}=\ket{\Omega}$ and all $z_i=it_i$ imaginary we obtain the out-of-time-ordered modular multi-point correlators. It would be interesting to search for potential connections between these out-of-time-ordered correlators and the notions of modular chaos previously introduced in the literature \cite{boer2020holographic,ceyhan2020recovering}.

It is important to note that in our definition of the multi-state R\'enyi divergences in (\ref{multiRenyi}) we restricted to the range $0\leq \theta_1+\cdots \theta_n\leq 1$ to make sure that the resulting vector is in $L^{2r}_\omega$. In principle, we can extend beyond this range, for instance, by making some $\theta_i$ negative. While the resulting multi-state measure would not always be finite, in an infinite dimensional system that is hyperfinite (approximated by matrix algebras arbitrarily well) one expects that this measure is finite for a large class of states $\psi_1, \cdots, \psi_n$. It would be interesting to explore the data processing inequality in this extended range.\footnote{Note that our proof only works when all $\theta_i$ are positive.} 

Finally, the analysis non-commutative $L^p_\omega$ spaces suggests that one might be able to prove an improved data processing inequality using Hirschman's lemma in the spirit of \cite{wilde2015recoverability,junge2020universal,faulkner2020approximate}. We postpone this to future work.

\paragraph{Acknowledgements:}

We thank Roy Araiza, Stefan Hollands, Nicholas LaRacuente and Thomas Sinclair for insightful discussions on non-commutative $L^p$ spaces. We also thank Milan Mosonyi and Mark Wilde for comments on the draft. NL is very grateful to the DOE that supported this work through grant DE-SC0007884 and the QuantiSED Fermilab consortium.
% This work was supported, in part, by a grant-in-aid (PHY1911298) from the National Science Foundation.

\appendix

\section{Riesz-Thorin theorem}\label{App:Riezs}

In this appendix, we prove the Riesz-Thorin theorem for the Araki-Masuda $(p,\omega)$-norms \cite{berta2018renyi}. Consider the algebras $\mA$ and $\mB$, faithful states $\omega_A$ and $\omega_B$ and their corresponding GNS Hilbert space $\mH_{A}$ and $\mH_{B}$, respectively. For a bounded linear map $T:\mH_{A}\to \mH_{B}$ and $p,q\geq 2$ as in (\ref{supernorm}) and (\ref{supernorm2}) we define the $(p,A)\to (q,B)$ norm to be
\begin{eqnarray}\label{pqNorm}
   \|T\|_{(p,A)\to (q,B)}=\sup_{\ket{\chi}\in \mH_A}\frac{\|T \ket{\chi}\|_{(q,\Omega_B)}}{\|\ket{\chi}\|_{(p,\Omega_A)}}\ .
\end{eqnarray}
Then, for 
\begin{eqnarray}
   &&\frac{1}{p_\theta}=\frac{1-\theta}{p_0}+\frac{\theta}{p_1}\nn\\
   &&\frac{1}{q_\theta}=\frac{1-\theta}{q_0}+\frac{\theta}{q_1},
\end{eqnarray}
we have the inequality 
\begin{eqnarray}
   \|T\|_{(p_\theta,A)\to (q_\theta,B)}\leq \|T\|_{(p_0,A)\to (q_0,B)}^{1-\theta}\|T\|_{(p_1,A)\to (q_1,B)}^\theta\ .
\end{eqnarray}
To prove this inequality, we first use the fact that any $\|\ket{\chi}\|_{p_\theta,A}=1$ can be written as $u\Delta_{\chi|\Omega;A}^{1/p_\theta}\ket{\omega_A}$ to write the left-hand-side as
\begin{eqnarray}
 \|T\|_{(p_\theta,A)\to (q_\theta,B)}=\sup_{\ket{\chi}\in \mH_A,u\in \mA}\|Tu\Delta_{\chi|\Omega;A}^{1/p_\theta}\ket{\Omega_A}\|_{q_\theta,\Omega_B}\ .
 \end{eqnarray}
We can use the definition of the $(q_\theta,\Omega_B)$ norm in (\ref{arakimasuda2}) to write the expression above as
\begin{eqnarray}
 \|T\|_{(p_\theta,A)\to (q_\theta,B)}=\sup_{u\in \mA,\ket{\chi}\in \mH_A, \ket{\Phi}\in \mH_B}\|\Delta_{\Phi|\Omega;B}^{\frac{1}{2}-\frac{1}{q_\theta}}T u\Delta_{\chi|\Omega;A}^{\frac{1}{p_\theta}}\ket{\Omega_A}\|\ .
\end{eqnarray}
We define the function 
\begin{eqnarray}
 f(\theta)=\|\Delta_{\Phi|\Omega;B}^{\frac{1}{2}-\frac{1}{q_\theta}}T u\Delta_{\chi|\Omega;A}^{\frac{1}{p_\theta}}\ket{\Omega_A}\|
\end{eqnarray}
and then analytically continue $\theta\to z=\theta+i t$ to the complex strip $\theta\in [0,1]$. This function is bounded, holomorphic everywhere inside the strip and is continuous on the boundaries of the strip at $\theta=1$ and $\theta=0$. Therefore, by the Phragm\'en-Lindel\"{o}f principle, it takes its maximum value on the boundaries of the strip. Using the Hadamard three line theorem, we find 
\begin{eqnarray}
 \|\Delta_{\Phi|\Omega;B}^{\frac{1}{2}-\frac{1}{q_\theta}}Tu\Delta_{\chi|\Omega;A}^{\frac{1}{p_\theta}}\ket{\Omega_A}\|\leq \|\Delta_{\Phi|\Omega;B}^{\frac{1}{2}-\frac{1}{q_0}}Tu \Delta_{\chi|\Omega;A}^{\frac{1}{p_0}}\ket{\Omega_A}\|^{(1-\theta)}\|\Delta_{\Phi|\Omega;B}^{\frac{1}{2}-\frac{1}{q_1}}Tu \Delta_{\chi|\Omega;A}^{\frac{1}{p_1}}\ket{\Omega_A}\|^{\theta}\ .\nn
\end{eqnarray}
Taking the supremum of both sides and using $\sup(fg)\leq \sup(f)\sup(g)$ implies the proof
\begin{eqnarray}
 \|T\|_{(p_\theta,A)\to (q_\theta,B)}\leq \|T\|_{(p_0,A)\to (q_0,B)}^{1-\theta}\|T\|_{(p_1,A)\to (q_1,B)}^\theta\ .
\end{eqnarray}

\section{Extended range of \texorpdfstring{$\theta$}{}}\label{app:extended}

Consider the $(\theta,r)$-R\'enyi divergence. If we choose $\theta\in (-1/2,0]$ the measure need not be finite. However, for a dense set of states it is finite. To see this, first assume that there exists a positive constant $c$ such that for all $a_+\in \mA$ we have
\begin{eqnarray}\label{majorize}
 \omega(a_+)\leq c \psi(a_+)\ .
\end{eqnarray}
In the density matrix setting, it means that the following operator is positive semi-definite
\begin{eqnarray}
 c\psi-\omega\geq 0\ .
\end{eqnarray}
Since the map $\Phi^*$ is CP we also have
\begin{eqnarray}
 c\Phi^*(\psi)-\Phi^*(\omega)\geq 0\ .
\end{eqnarray}
For such states we have 
\begin{eqnarray}
 \braket{a\Omega|\Delta_{\Psi|\Omega}a\Omega}=\braket{a^\dagger\Psi|a^\dagger\Psi}\geq c^{-1}\braket{a^\dagger\Omega|a^\dagger \Omega}=c^{-1}\braket{a\Omega|\Delta_\Omega a\Omega}
\end{eqnarray}
which implies the inequality $c\Delta_{\Psi|\Omega}\geq \Delta_\Omega$. 
For $\theta\in [0,1]$ we obtain\footnote{See also Lemma 5 of \cite{faulkner2020approximate}.}
\begin{eqnarray}
 c^\theta\geq \Delta_\Omega^{\theta/2}\Delta_{\Psi|\Omega}^{-\theta}\Delta_\Omega^{\theta/2}\ .
\end{eqnarray}
This implies
\begin{eqnarray}
 c^\theta\geq \|\Delta_{\Psi|\Omega}^{-\theta/2}\Delta_\Omega^{\theta/2}\|_{\infty,\Omega}\ .
\end{eqnarray}
Therefore, the condition in (\ref{majorize}) says that the vector
\begin{eqnarray}
 \Delta^{\theta/2}_{\Psi|\Omega}\ket{\Omega}\in \mH_\Omega
\end{eqnarray}
for $\theta\in [-1,1]$. For $r\geq 1$ this vector is in $L^{2r}_\omega$, therefore
\begin{eqnarray}
 S_{\theta,r}(\psi\|\Omega)=\frac{-2r}{1-\theta}\log\|\Delta^{\theta/2}_{\Psi|\Omega}\ket{\Omega}\|_{2r,\Omega}
\end{eqnarray}
is finite. 

\section{The relative entropy limit}\label{app:theta}

This appendix uses arguments similar to those in \cite{faulkner2020approximate}.
Consider the family of vectors $\ket{\chi_\epsilon}\in L^{2r}_\omega$ such that $\ket{\chi_\epsilon}=\ket{\Omega}+\epsilon\ket{\chi_1}+O(\epsilon^2)$. If we normalize the vector $\ket{\chi_\epsilon}$ to $\ket{\bar{\chi}_\epsilon}=\ket{\chi_\epsilon}/\|\ket{\chi_\epsilon}\|_{2,\Omega}$ we obtain
\begin{eqnarray}\label{limitepsilon}
 \lim_{\epsilon\to 0}\frac{1}{2\epsilon}\|\ket{\bar{\chi}_\epsilon}-\ket{\Omega}\|^2=\lim_{\epsilon\to 0}\frac{1}{\epsilon}\lb 1-\Re\braket{\bar{\chi}_\epsilon|\Omega}\rb=0\ .
\end{eqnarray}
% \SO{We can show that the real part goes to 1 since $2\lim_{\epsilon\to 0}\Re\braket{\bar{\chi}_\epsilon|\Omega} = \lim_{\epsilon\to 0}(\braket{\bar{\chi}_\epsilon|\Omega}+\braket{\Omega|\bar{\chi}_\epsilon}) = 2$. }
Next, we note that for $r\geq 1$ we have 
\begin{eqnarray}\label{ineqchi}
 \Re\braket{\bar{\chi}_\epsilon|\Omega}\leq |\braket{\bar{\chi}_\epsilon|\Omega}|\leq \|\ket{\bar{\chi}_\epsilon}\|_{2r,\Omega}\|\ket{\Omega}\|_{s,\Omega}=\|\ket{\bar{\chi}_\epsilon}\|_{2r,\Omega}\leq \|\ket{\bar{\chi}_\epsilon}\|^{1/r}_{2,\Omega}=1
\end{eqnarray}
where in the second inequality we have used the H\"{o}lder inequality and the fact that the $(s,\Omega)$-norm of $\ket{\Omega}$ is always one. In the last inequality, we have used the fact that for $r\geq p$
\begin{eqnarray}
   \|\ket{\Psi}\|_{r,\Omega}^{r}\leq \|\ket{\Psi}\|_{p,\Omega}^p\ .
\end{eqnarray}
This follows from a simple application of the Hadamard three-line theorem to the function $\|\ket{\Psi}\|_{r,\Omega}^r$; see lemma 8 and corollary 5 of \cite{berta2018renyi} for more detail.

Divide (\ref{ineqchi}) by $\epsilon$ and take the limit $\epsilon\to 0$. Using (\ref{limitepsilon}) we obtain
\begin{eqnarray}
 \lim_{\epsilon\to 0}\frac{1}{\epsilon}\lb 1-\|\ket{\bar{\chi}_\epsilon}\|_{2r,\Omega}\rb=0\ .
\end{eqnarray}
As a result,
\begin{eqnarray}
 \lim_{\epsilon\to 0}\frac{1}{\epsilon}\log \|\ket{\bar{\chi}_\epsilon}\|_{2r,\Omega}=\partial_\epsilon \lb\|\ket{\bar{\chi}_\epsilon}\|_{2r,\Omega}\rb_{\epsilon\to 0}=0\ .
\end{eqnarray}
We are interested in the function
\begin{eqnarray}
 \lim_{\epsilon\to 0}\frac{1}{\epsilon}\log\|\ket{\chi_\epsilon}\|_{2r,\Omega}\ .
\end{eqnarray}
The $(p,\Omega)$-norms are homogeneous therefore
\begin{eqnarray}
 \log\|\ket{\chi_\epsilon}\|_{2r,\Omega}=\log\|\ket{\bar{\chi}_\epsilon}\|_{2r,\Omega}+\log \|\ket{\chi_\epsilon}\|_{2,\Omega}
\end{eqnarray}
and 
\begin{eqnarray}
 \lim_{\epsilon\to 0}\frac{1}{\epsilon}\log \|\ket{\chi_\epsilon}\|_{2r,\Omega}=\lim_{\epsilon\to 0}\frac{1}{\epsilon}\log \|\ket{\chi_\epsilon}\|_{2,\Omega}\ .
\end{eqnarray}
Therefore, we only need to study the $(2,\Omega)$-norm of the vector $\ket{\chi_\epsilon}$.

In the three-state R\'enyi measures our vector of interest is
\begin{eqnarray}
 \ket{\chi_\epsilon}=\lb\Delta_{\Psi_1|\Omega}^{\frac{\epsilon \beta}{(1-\alpha)r}}\sharp_\alpha \Delta_{\Psi_2|\Omega}^{\frac{\epsilon (1-\beta)}{\alpha r}}\rb^{1/2}\ket{\Omega}\ .
\end{eqnarray}
We have
\begin{eqnarray}\label{derivativeeps}
 \lim_{\epsilon\to 0}\frac{1}{\epsilon}\log\|\ket{\chi_\epsilon}\|_{2r,\Omega}=\frac{1}{2}\braket{\Omega|\partial_{\epsilon}\lb \Delta_{\Psi_1|\Omega}^{\frac{\epsilon \beta}{(1-\alpha)r}}\sharp_\alpha \Delta_{\Psi_2|\Omega}^{\frac{\epsilon (1-\beta)}{\alpha r}}\rb_{\epsilon\to 0}|\Omega}\ .
\end{eqnarray}
We only need to compute the derivative:
\begin{eqnarray}
 \partial_{\epsilon}\lb X^{\epsilon}\sharp_\alpha Y^\epsilon\rb_{\epsilon\to 0}&&=\log X+\partial_\epsilon\lb X^{-\epsilon/2}Y^\epsilon X^{-\epsilon/2}\rb^\alpha\Big|_{\epsilon=0}\nn\\
 &&=(1-\alpha)\log X+\alpha \log Y\ .
\end{eqnarray}
Applied to our case in (\ref{derivativeeps}) we find
\begin{eqnarray}
  \lim_{\epsilon\to 0}\frac{1}{\epsilon}\log\|\ket{\chi_\epsilon}\|_{2r,\Omega}=\frac{-1}{2r}\lb\beta S(\psi_1\|\omega)+(1-\beta)S(\psi_2\|\omega)\rb\ .
\end{eqnarray}
As a result, from eq (\ref{moresymm}) we get
\begin{eqnarray}
  \lim_{\epsilon \to 0} \hat{S}_{(\epsilon \beta, \epsilon(1-\beta)),r}^\alpha(\psi_1,\psi_2\|\omega)=\beta S(\psi_1\|\omega)+(1-\beta)S(\psi_2\|\omega)\ .
\end{eqnarray}
To generalize to $n$ states we need to compute
\begin{eqnarray}
   \partial_{\epsilon}\lb X_1^{\epsilon}\sharp_{\alpha_1} \cdots \sharp_{\alpha_{n-1}}X_n^\epsilon\rb_{\epsilon\to 0}
&&=(1-\alpha_1)\log X_1+\alpha_1 \partial_{\epsilon}\lb X_2^{\epsilon}\sharp_{\alpha_2}\cdots \sharp_{\alpha_{n-1}}X_n^\epsilon \rb_{\epsilon\to 0}\nn\\
&&=\gamma_1\log X_1+\gamma_2 \log X_2+\cdots +\gamma_n \log X_n\ .
\end{eqnarray}
Consider the vector
\begin{eqnarray}
   \ket{\chi_\epsilon}=\lb \Delta_{\Psi_1|\Omega}^{\frac{\epsilon \beta_1}{\gamma_1 r}}\sharp_{\alpha_1}\cdots \sharp_{\alpha_{n-1}}\Delta_{\Psi_n|\Omega}^{\frac{\epsilon \beta_n}{\gamma_n r}}\rb^{1/2}\ket{\Omega}\ .
\end{eqnarray}
Then,
\begin{eqnarray}
   \lim_{\epsilon\to 0}\hat{S}^{\vec{\alpha}}_{\vec{\theta}_\epsilon,r}(\vec{\psi}\|\omega)=\sum_{i=1}^n \beta_i S(\psi_i\|\omega)\ .
\end{eqnarray}

\section{The $(p\to q)$-norm of contractions}\label{app:proof}

Consider a linear operator $F_f : \mH_B \to \mH_A$ that satisfies $\| F_f^{\dagger} F_f \|_{\infty} \leq 1$; see equation (\ref{FfFf}).
We prove that for $\forall p\in[2,\infty]$
\begin{equation}\label{toprove}
    \|F_f\|_{(p,\Omega_B) \to (p,\Omega_A)} \leq 1\ .
\end{equation}
% We show the stronger statement that  $\|F_f\|_{(p,\Omega_B) \to (p,\Omega_A)} \leq 1$ for 

{\bf Proof:} First, note that $\| F_f^{\dagger} F_f \|_{\infty} \leq 1$ implies that $F_f$ is a contraction, i.e. $\| F_f \|_{\infty} \leq 1$, because $\| T^{\dagger}T \|_{p} = \| T \|_{2p}^2$ for any linear operator $T:\mH_B \to \mH_A$ and $\forall p \in[1,\infty]$. 
The proof has two steps: First, we show that for a contraction $F_f$ we have $ \|F_f \|_{(p,\Omega_B) \to (p,\Omega_A) }  \leq \| F_f \|_{\infty} $ for $p=2, \infty$. Then, we use the Riesz-Thorin interpolation theorem to establish \ref{toprove}.

For the first step, consider an isometry $V:\mH_B \hookrightarrow \mH_A$ and a cyclic and separating vector $\ket{\Omega_B} = V^{\dagger} \ket{\Omega_A}$.
% and  $VV^{\dagger} = P_A$ is a projection on $\mH_A$ that defines the subspace $V\mH_B = P_A \mH_A \subset \mH_A $. 
For $p=2$ we have
\begin{equation}
    \begin{split}
         \|F_f \|_{(2, \Omega_B)\to (2, \Omega_A)} :=& \underset{b\in \mB}{\sup} \frac{\|F_f b\ket{\Omega_B} \|_{2,\Omega_A}}{\| b\ket{\Omega_B} \|_{2,\Omega_B} } \\
         =& \underset{b\in \mB}{\sup}  \frac{\|F_f V^{\dagger} V b V^{\dagger}\ket{\Omega_A} \|_{2,\Omega_A}}{\| b\ket{\Omega_B} \|_{2,\Omega_B} } \\
         \leq&   \|F_f V^{\dagger} \|_{\infty}\, \underset{b\in \mB}{\sup}\frac{\| V b V^{\dagger}\ket{\Omega_A} \|_2}{\| b\ket{\Omega_B} \|_{2} } \\
         \leq & \|F_f V^{\dagger}\|_{\infty} \\
         \leq & \|F_f \|_{\infty}\ .
    \end{split}
\end{equation}
In the third line, we have used H\"older's inequality and (\ref{2omega}). By a similar argument, for $p=\infty$, we obtain
\begin{equation}
    \begin{split}
         \|F_f \|_{(\infty, \Omega_B)\to (\infty, \Omega_A)} :=& \underset{b\in \mB}{\sup} \frac{\|F_f b\ket{\Omega_B} \|_{\infty,\Omega_A}}{\| b\ket{\Omega_B} \|_{\infty,\Omega_B} } \\
         =& \underset{b\in \mB}{\sup}  \frac{\|F_f V^{\dagger} V b V^{\dagger}\ket{\Omega_A} \|_{\infty,\Omega_A}}{\| b\ket{\Omega_B} \|_{\infty,\Omega_B} } \\
         \leq&   \|F_f V^{\dagger} \|_{\infty}\, \underset{b\in \mB}{\sup}\frac{\| V b V^{\dagger}\|_{\infty}}{\| b\|_{\infty} } \\
         \leq & \|F_f V^{\dagger}\|_{\infty} \\
         \leq & \|F_f \|_{\infty}\ .
    \end{split}
\end{equation}
where we have used $\|a\ket{\Omega_A}\|_{\infty,\Omega_A}=\|a\|_\infty$.
% The first step in both inequalities above is the definition of $\|\cdot \|_{(p, \Omega_B)\to (p, \Omega_A)}$. 
% To proceed to the next equality, we notice that
% $\| V b V^{\dagger} \ket{\Omega_A}\|_p= \| V b \ket{\Omega_B}\|_p \leq \|b \ket{\Omega_B}\|_p$ since $V$ is an isometry (for $p=2$ we have an equality). The last inequality follows from $\|F_f V^{\dagger}\|_{\infty} \leq \|F_f \|_{\infty} \| V^{\dagger}\|_{\infty} \leq 1$. 
Since $\| F_f\|_{\infty} \leq 1$, the above inequalities imply that for $p=2$ or $p=\infty$
\begin{equation}\label{eq:p-pcontract}
    \|F_f \|_{(p, \Omega_B)\to (p, \Omega_A)} \leq 1
\end{equation}

In the second step, we use the Riesz-Thorin interpolation theorem,
\begin{equation}
    \|F_f\|_{(p_{\theta},\Omega_B)\to (p_{\theta},\Omega_A)} \leq \|F_f \|^{1-\theta}_{(\infty,\Omega_B)\to \infty,\Omega_A)} \|F_f \|^{\theta}_{(2,\Omega_B)\to 2,\Omega_A)}
\end{equation}
for $\frac{1}{p_{\theta}} = \frac{1-\theta}{p_0} + \frac{\theta}{p_1}$ with $\theta \in [0,1]$ where we set $p_0 = \infty$ and $p_1 = 2$. From equation (\ref{eq:p-pcontract}),
\begin{equation}
    \|F_f\|_{(p_{\theta},\Omega_B)\to p_{\theta},\Omega_A)} \leq 1
\end{equation}
for $\forall p_{\theta}\in [2,\infty]$. Just by relabeling $p_\theta$ to $p$, we obtain the statement in (\ref{toprove}).$\Box$

%\section{Tentative: Hierarchy of norms}

%For an isometry $v:\mH_A \to \mH_A$, i.e. $\|v \|_{\infty} \leq 1$, one can show $\| v \|_{(\infty, \omega_A)  \to (\infty, \omega_A) } \leq 1$. This proof is based on the following fact. For a linear operator $T : \mH_A \to \mH_A$, 
%\begin{equation}
%    \| T \|_{\infty} = \| T \|_{(\infty, \omega)} \geq\| T \|_{(p, \omega) \to (p, \omega)}
%\end{equation}
%for any $p \in [1, \infty]$ and a state $\omega$. The above inequalities can be shown by their definition and Holder's inequality. When $T$ is an contraction, i.e. $\| T\|_{\infty} \leq 1$, then, it is clear from the above that $\| T\|_{(\infty, \omega) \to (\infty, \omega)} \leq 1$.

\bibliographystyle{JHEP}
\bibliography{main}

\providecommand{\href}[2]{#2}\begingroup\raggedright\begin{thebibliography}{10}

\bibitem{matsumoto2015new}
K.~Matsumoto, {\it A new quantum version of f-divergence},  in {\em Nagoya
  Winter Workshop: Reality and Measurement in Algebraic Quantum Theory},
  pp.~229--273, Springer, 2015.

\bibitem{vedral2002role}
V.~Vedral, {\it The role of relative entropy in quantum information theory},
  {\em Reviews of Modern Physics} {\bf 74} (2002), no.~1 197.

\bibitem{hiai1991proper}
F.~Hiai and D.~Petz, {\it The proper formula for relative entropy and its
  asymptotics in quantum probability},  {\em Communications in mathematical
  physics} {\bf 143} (1991), no.~1 99--114.

\bibitem{mosonyi2015quantum}
M.~Mosonyi and T.~Ogawa, {\it {Quantum hypothesis testing and the operational
  interpretation of the quantum R{\'e}nyi relative entropies}},  {\em
  Communications in Mathematical Physics} {\bf 334} (2015), no.~3 1617--1648.

\bibitem{audenaert2015alpha}
K.~M. Audenaert and N.~Datta, {\it {$\alpha$-z-R{\'e}nyi relative entropies}},
  {\em Journal of Mathematical Physics} {\bf 56} (2015), no.~2 022202.

\bibitem{jaksic2011entropic}
V.~Jaksic, Y.~Ogata, Y.~Pautrat, and C.-A. Pillet, {\it {Entropic fluctuations
  in quantum statistical mechanics. An introduction}},  {\em arXiv preprint
  arXiv:1106.3786} (2011).

\bibitem{zhang2020wigner}
H.~Zhang, {\it {From Wigner-Yanase-Dyson conjecture to Carlen-Frank-Lieb
  conjecture}},  {\em Advances in Mathematics} {\bf 365} (2020) 107053.

\bibitem{brandao2010generalization}
F.~G. Brandao and M.~B. Plenio, {\it {A generalization of quantum Stein’s
  lemma}},  {\em Communications in Mathematical Physics} {\bf 295} (2010),
  no.~3 791--828.

\bibitem{araki1982positive}
H.~Araki and T.~Masuda, {\it {Positive cones and Lp-spaces for von Neumann
  algebras}},  {\em Publications of the Research Institute for Mathematical
  Sciences} {\bf 18} (1982), no.~2 759--831.

\bibitem{hiai2013concavity}
F.~Hiai, {\it Concavity of certain matrix trace and norm functions},  {\em
  Linear algebra and its applications} {\bf 439} (2013), no.~5 1568--1589.

\bibitem{Wilde_2015}
M.~M. Wilde, {\it Multipartite quantum correlations and local recoverability},
  {\em Proceedings of the Royal Society A: Mathematical, Physical and
  Engineering Sciences} {\bf 471} (May, 2015) 20140941.

\bibitem{dupuis2016swiveled}
F.~Dupuis and M.~M. Wilde, {\it {Swiveled R{\'e}nyi entropies}},  {\em Quantum
  Information Processing} {\bf 15} (2016), no.~3 1309--1345.

\bibitem{berta2015renyi}
M.~Berta, K.~P. Seshadreesan, and M.~M. Wilde, {\it R{\'e}nyi generalizations
  of quantum information measures},  {\em Physical Review A} {\bf 91} (2015),
  no.~2 022333.

\bibitem{beigi2013sandwiched}
S.~Beigi, {\it {Sandwiched R{\'e}nyi divergence satisfies data processing
  inequality}},  {\em Journal of Mathematical Physics} {\bf 54} (2013), no.~12
  122202.

\bibitem{kosaki1984applications}
H.~Kosaki, {\it {Applications of the complex interpolation method to a von
  Neumann algebra: non-commutative L$_p$-spaces}},  {\em Journal of functional
  analysis} {\bf 56} (1984), no.~1 29--78.

\bibitem{wilde2014strong}
M.~M. Wilde, A.~Winter, and D.~Yang, {\it {Strong converse for the classical
  capacity of entanglement-breaking and Hadamard channels via a sandwiched
  R{\'e}nyi relative entropy}},  {\em Communications in Mathematical Physics}
  {\bf 331} (2014), no.~2 593--622.

\bibitem{muller2013quantum}
M.~M{\"u}ller-Lennert, F.~Dupuis, O.~Szehr, S.~Fehr, and M.~Tomamichel, {\it
  {On quantum R{\'e}nyi entropies: A new generalization and some properties}},
  {\em Journal of Mathematical Physics} {\bf 54} (2013), no.~12 122203.

\bibitem{frank2013monotonicity}
R.~L. Frank and E.~H. Lieb, {\it Monotonicity of a relative r{\'e}nyi entropy},
   {\em Journal of Mathematical Physics} {\bf 54} (2013), no.~12 122201.

\bibitem{lashkari2019constraining}
N.~Lashkari, {\it Constraining quantum fields using modular theory},  {\em
  Journal of High Energy Physics} {\bf 2019} (2019), no.~1 59.

\bibitem{wilde2018optimized}
M.~M. Wilde, {\it Optimized quantum f-divergences and data processing},  {\em
  Journal of Physics A: Mathematical and Theoretical} {\bf 51} (2018), no.~37
  374002.

\bibitem{petz1985quasi}
D.~Petz, {\it {Quasi-entropies for states of a von Neumann algebra}},  {\em
  Publications of the Research Institute for Mathematical Sciences} {\bf 21}
  (1985), no.~4 787--800.

\bibitem{petz1986quasi}
D.~Petz, {\it Quasi-entropies for finite quantum systems},  {\em Reports on
  mathematical physics} {\bf 23} (1986), no.~1 57--65.

\bibitem{petz2003monotonicity}
D.~Petz, {\it Monotonicity of quantum relative entropy revisited},  {\em
  Reviews in Mathematical Physics} {\bf 15} (2003), no.~01 79--91.

\bibitem{witten2018aps}
E.~Witten, {\it Aps medal for exceptional achievement in research: Invited
  article on entanglement properties of quantum field theory},  {\em Reviews of
  Modern Physics} {\bf 90} (2018), no.~4 045003.

\bibitem{schilling2012bernstein}
R.~L. Schilling, R.~Song, and Z.~Vondracek, {\em Bernstein functions: theory
  and applications}, vol.~37.
\newblock Walter de Gruyter, 2012.

\bibitem{bhatia2013matrix}
R.~Bhatia, {\em Matrix analysis}, vol.~169.
\newblock Springer Science \& Business Media, 2013.

\bibitem{faulkner2020approximate}
T.~Faulkner and S.~Hollands, {\it Approximate recoverability and relative
  entropy ii: 2-positive channels of general v. neumann algebras},  {\em arXiv
  preprint arXiv:2010.05513} (2020).

\bibitem{kubo1980means}
F.~Kubo and T.~Ando, {\it Means of positive linear operators},  {\em
  Mathematische Annalen} {\bf 246} (1980), no.~3 205--224.

\bibitem{simon2019operator}
B.~Simon, {\it {Operator Means, II: Kubo--Ando Theorem}},  in {\em Loewner's
  Theorem on Monotone Matrix Functions}, pp.~379--384.
\newblock Springer, 2019.

\bibitem{sagae1994upper}
M.~Sagae and K.~Tanabe, {\it Upper and lower bounds for the
  arithmetic-geometric-harmonic means of positive definite matrices},  {\em
  Linear and Multilinear Algebra} {\bf 37} (1994), no.~4 279--282.

\bibitem{sutter2017multivariate}
D.~Sutter, M.~Berta, and M.~Tomamichel, {\it Multivariate trace inequalities},
  {\em Communications in Mathematical Physics} {\bf 352} (2017), no.~1 37--58.

\bibitem{hiai1993golden}
F.~Hiai and D.~Petz, {\it {The Golden-Thompson trace inequality is
  complemented}},  {\em Linear algebra and its applications} {\bf 181} (1993)
  153--185.

\bibitem{ahn2007extended}
E.~Ahn, S.~Kim, and Y.~Lim, {\it {An extended Lie--Trotter formula and its
  applications}},  {\em Linear algebra and its applications} {\bf 427} (2007),
  no.~2-3 190--196.

\bibitem{haag2012local}
R.~Haag, {\em Local quantum physics: Fields, particles, algebras}.
\newblock Springer Science \& Business Media, 2012.

\bibitem{connes1973classification}
A.~Connes, {\it {A classification of factors of type III}},  in {\em Scientific
  Annals of the {\ 'E} cole Normale Sup {\' e} rieure}, vol.~6, pp.~133--252,
  1973.

\bibitem{audenaert2007discriminating}
K.~M. Audenaert, J.~Calsamiglia, R.~Munoz-Tapia, E.~Bagan, L.~Masanes, A.~Acin,
  and F.~Verstraete, {\it {Discriminating states: The quantum Chernoff bound}},
   {\em Physical review letters} {\bf 98} (2007), no.~16 160501.

\bibitem{bjelakovic2005quantum}
I.~Bjelakovi{\'c}, J.-D. Deuschel, T.~Kr{\"u}ger, R.~Seiler,
  R.~Siegmund-Schultze, and A.~Szko{\l}a, {\it {A quantum version of Sanov's
  theorem}},  {\em Communications in mathematical physics} {\bf 260} (2005),
  no.~3 659--671.

\bibitem{hayashi2002optimal}
M.~Hayashi, {\it {Optimal sequence of quantum measurements in the sense of
  Stein's lemma in quantum hypothesis testing}},  {\em Journal of Physics A:
  Mathematical and General} {\bf 35} (2002), no.~50 10759.

\bibitem{li2016discriminating}
K.~Li et~al., {\it {Discriminating quantum states: The multiple Chernoff
  distance}},  {\em Annals of Statistics} {\bf 44} (2016), no.~4 1661--1679.

\bibitem{mosonyi2020error}
M.~Mosonyi, Z.~Szil{\'a}gyi, and M.~Weiner, {\it On the error exponents of
  binary quantum state discrimination with composite hypotheses},  {\em arXiv
  preprint arXiv:2011.04645} (2020).

\bibitem{brandao2020adversarial}
F.~G. Brandao, A.~W. Harrow, J.~R. Lee, and Y.~Peres, {\it {Adversarial
  hypothesis testing and a quantum Stein’s lemma for restricted
  measurements}},  {\em IEEE Transactions on Information Theory} {\bf 66}
  (2020), no.~8 5037--5054.

\bibitem{fawzi2021defining}
H.~Fawzi and O.~Fawzi, {\it Defining quantum divergences via convex
  optimization},  {\em Quantum} {\bf 5} (2021) 387.

\bibitem{boer2020holographic}
J.~Boer and L.~Lamprou, {\it Holographic order from modular chaos},  {\em
  Journal of High Energy Physics} {\bf 2020} (2020), no.~1912.02810 1--24.

\bibitem{ceyhan2020recovering}
F.~Ceyhan and T.~Faulkner, {\it {Recovering the QNEC from the ANEC}},  {\em
  Communications in Mathematical Physics} {\bf 377} (2020), no.~2 999--1045.

\bibitem{wilde2015recoverability}
M.~M. Wilde, {\it Recoverability in quantum information theory},  {\em
  Proceedings of the Royal Society A: Mathematical, Physical and Engineering
  Sciences} {\bf 471} (2015), no.~2182 20150338.

\bibitem{junge2020universal}
M.~Junge and N.~LaRacuente, {\it Universal recovery and p-fidelity in von
  neumann algebras},  {\em arXiv preprint arXiv:2009.11866} (2020).

\bibitem{berta2018renyi}
M.~Berta, V.~B. Scholz, and M.~Tomamichel, {\it {R{\'e}nyi Divergences as
  Weighted Non-commutative Vector-Valued $L_p$-Spaces}},  in {\em Annales Henri
  Poincar{\'e}}, vol.~19, pp.~1843--1867, Springer, 2018.

\end{thebibliography}\endgroup

\end{document}